\documentclass[sigconf, screen]{acmart}
\usepackage{multirow}
\usepackage{makecell}
\usepackage{graphicx}     
\usepackage[ruled,linesnumbered]{algorithm2e}
\usepackage{caption}      
\usepackage{pifont}
\usepackage{subcaption}
\usepackage{adjustbox}
\usepackage[table]{xcolor} 
\usepackage{booktabs}

\AtBeginDocument{%
  }

\setcopyright{none} 
\copyrightyear{2026}
\acmYear{2026}
\acmDOI{XXXXXXX.XXXXXXX}

\acmConference[TBD]{To Be Decided}{TBD}{TBD}
\acmISBN{000-0-0000-0000-0/00/00}

\begin{document}

\title{Do We Need Tensor Cores for Stencil Computations?}

 \author{Qiqi Gu}
\authornote{Equal Contribution}
\email{qiqi.gu@sjtu.edu.cn}
\affiliation{%
  \institution{Shanghai Jiao Tong University}
  \state{Shanghai}
  \country{China}
}

\author{Chenpeng Wu}
\authornotemark[1]
\email{cpwu_sjtu@sjtu.edu.cn}
\affiliation{%
  \institution{Shanghai Jiao Tong University}
  \state{Shanghai}
  \country{China}
}

\author{Heng Shi}
\email{heng.shi@sjtu.edu.cn}
\authornotemark[2]
\affiliation{%
  \institution{Shanghai Enflame Technology Co.,Ltd.; Shanghai Jiao Tong University}
  \state{Shanghai}
  \country{China}
}

\author{Jianguo Yao}
\email{jianguo.yao@sjtu.edu.cn}
\authornote{Corresponding author}
\affiliation{%
  \institution{Shanghai Jiao Tong University}
  \state{Shanghai}
  \country{China}
}

\author{Haibing Guan}
\email{hbguan@sjtu.edu.cn}
\affiliation{%
  \institution{Shanghai Jiao Tong University}
  \state{Shanghai}
  \country{China}
}


\begin{abstract}
Stencil computation constitutes a cornerstone of scientific computing, serving as a critical kernel in domains ranging from fluid dynamics to weather simulation. 
While stencil computations are conventionally regarded as memory-bound and thus unsuitable for compute-centric Tensor Cores, recent empirical studies have demonstrated significant speedups after applying Tensor Cores, forming an apparent contradiction. 

This paper resolves this contradiction by conducting a systematic performance analysis of stencil computations on Tensor Cores. We begin by revisiting the adaptation of stencils onto Tensor Cores, quantifying the computational redundancy introduced by the transformations required to satisfy hardware constraints. These metrics are subsequently integrated into an enhanced performance model that explicitly accounts for the arithmetic intensity shifts driven by temporal fusion. Guided by this formulation, we derive analytical criteria to determine the suitability of Tensor Cores for varying stencil workloads. By classifying operational regions, we identify the specific \textit{sweet spot} for effective acceleration and further demonstrate how Sparse Tensor Cores expand this profitable design space. Extensive evaluations on NVIDIA GPUs across SOTA implementations, including DRStencil, EBISU, ConvStencil, and SPIDER, validate our performance model and analytical criteria. These results demonstrate the effectiveness of our approach in guiding stencil performance optimization.

\end{abstract}





\maketitle

\section{Introduction}\label{sec.intro}

\begin{figure}[t]
  \includegraphics[width=0.5\linewidth]{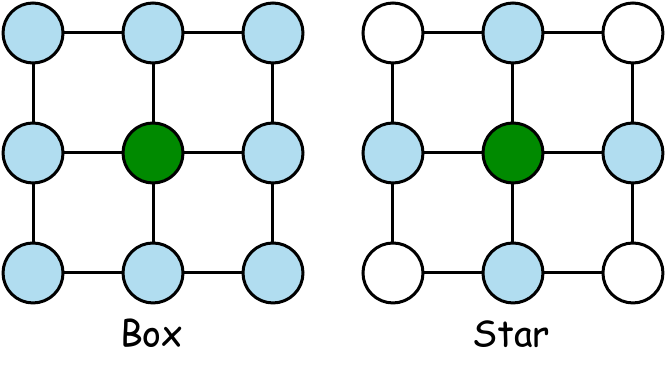} 
  \caption{Stencil Computation Pattern.}
  \label{fig.stencil}
\end{figure}

Stencil computation constitutes a foundational numerical method in which each grid point is updated via a weighted sum of its neighbors according to a fixed, pre-defined pattern. It serves as a cornerstone across diverse scientific domains, including fluid dynamics~\cite{huynh2014high, lusher2021opensbli}, earth system modeling~\cite{jacquelin2022scalable}, weather simulation~\cite{ao201726,ben2022productive}, and wave propagation~\cite{akbudak2020asynchronous, qu2023exploiting}, and is identified as one of the seven most critical numerical methods in science and engineering~\cite{pcolella, asanovic2006landscape}.
A stencil computation can be characterized by three key parameters: shape, radius \( r \) (also referred to as \textit{order}), and dimensionality \( d \). 
As shown in Figure~\ref{fig.stencil}, common shapes include the box stencil, which incorporates all grid points within a hyper-rectangular neighborhood centered on the target point, and the star stencil, which includes only points aligned with the coordinate axes.
The radius $r$ defines the neighborhood extent, that is, the maximum distance from the center point, while the dimensionality \( d \) specifies whether the domain is 1D, 2D, 3D, or higher.
For instance, the 2D Jacobi iterative method represents a typical example of a Star-2D1R stencil.

Conventionally, stencil computations are regarded as memory-bound workloads. This naturally stems from their workload pattern: a large number of memory accesses are required to fetch neighboring data points for each grid point update, while only a few computing operations are performed. For large-scale problems whose working sets exceed the capacity of the fast on-chip storage (e.g., registers or caches), frequent off-chip memory accesses become inevitable, placing significant pressure on memory bandwidth. Consequently, even though grid updates allow for massive parallelism, the system becomes memory-bound, leaving execution units idle.
Consequently, attainable performance is typically limited by the bandwidth of the memory subsystem, rather than the theoretical compute throughput of the hardware.

To release the memory pressure and better utilize the limited memory bandwidth, researchers have extensively analyzed the access pattern and proposed numerous optimizations that exploit spatial and temporal locality. Spatial tiling, being a representative optimization technique, partitions the computational domain into smaller blocks, enabling data to reside in faster memory where it can be reused across multiple computations. Specialized tiling schemes, including rectangular tiling~\cite{rivera2000tiling, nguyen20103}, cache-oblivious tiling~\cite{frigo2005cache, strzodka2010cache}, and tessellating tiling~\cite{yuan2017tessellating}, have been proposed to optimize stencil computations with a specific pattern. Another typical strategy is temporal fusion, it merges multiple time steps in a single computing kernel, exploiting inter-step locality to amortize memory access overhead. DRStencil~\cite{you2021drstencil} and EBISU~\cite{zhang2023revisiting} are two recent representatives that employ this optimization. Moreover, there exist some dedicated optimizations to better align with modern hardware features. For example, data layout transformations~\cite{jang2010data, henretty2011data, henretty2013stencil} reorganize memory representations to match hardware access patterns, reducing potential memory transaction amplification and resource contention, thus maximizing bandwidth utilization.

\begin{figure}[t]
  \includegraphics[width=.95\linewidth]{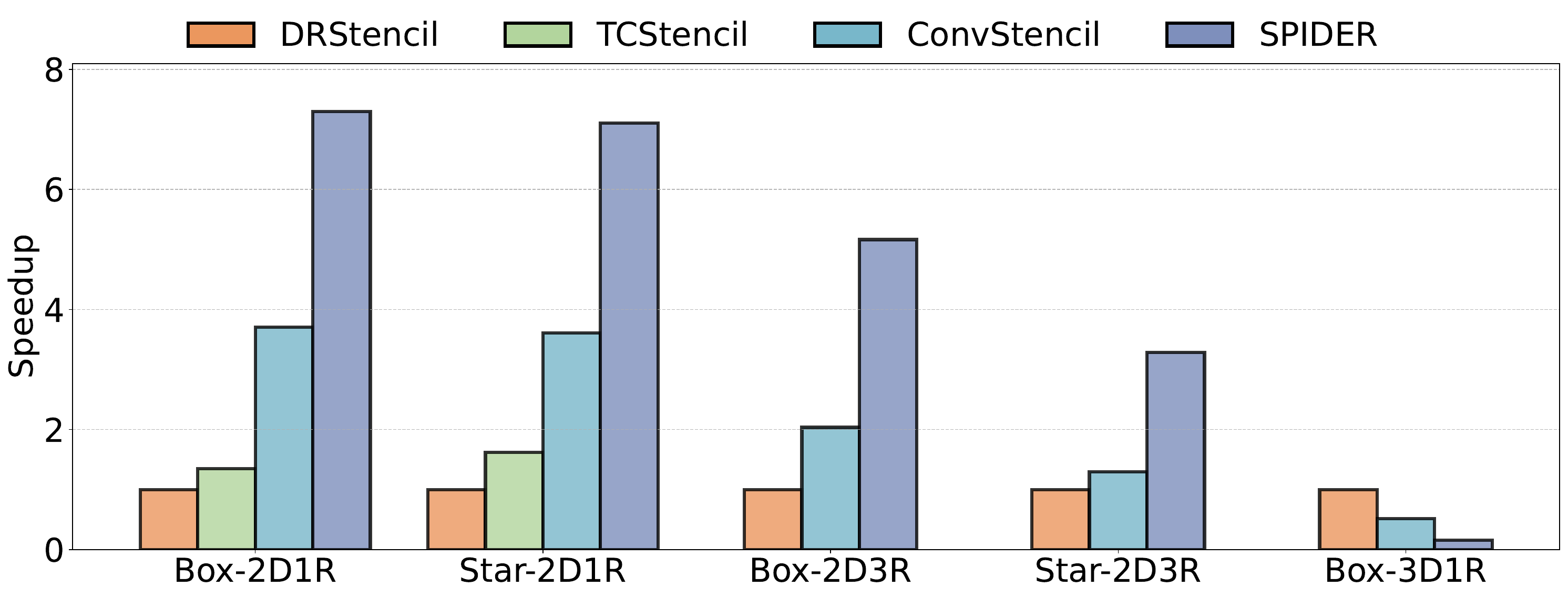} 
  \caption{Performance Comparison between CUDA Core and Tensor Core Implementations.}
  \label{fig.perf_compare}
\end{figure}

Given this inherent memory-bound nature of stencil computation, a long-standing consensus in the community is that emerging compute-centric Arithmetic Logic Units (ALUs), such as Tensor Cores and Matrix Cores, are not suitable for stencil workloads~\cite{CanTensorCores}. The rationale is straightforward: 
these units significantly increase compute throughput but offer no additional memory bandwidth compared to general-purpose ALUs (e.g., CUDA Cores), the performance of stencil kernels remains bounded by data movement. Therefore, simply adapting computations onto Tensor Cores theoretically yields negligible performance benefit over  CUDA Cores.

Interestingly, recent empirical evidence contradicts this conventional assumption. Several cutting-edge studies successfully adapt stencil computations onto Tensor Cores, achieving substantial speedups. For example, TCStencil~\cite{TCStencil} first demonstrates the feasibility of adapting star- and box-shaped stencils onto Tensor Cores. Subsequently, ConvStencil~\cite{chen2024convstencil} utilizes \textit{stencil2row} transformations and \textit{dual tessellation} to further optimize this adaptation, while SPIDER~\cite{SPIDER} introduces \textit{Strided Swapping} to further enable sparse ALUs with higher compute throughput. As illustrated in Figure~\ref{fig.perf_compare}, these approaches outperform CUDA Core implementation, DRStencil, by 1.48$\times$, 2.23$\times$, and 4.60$\times$, respectively. This presents a fundamental paradox: 
Tensor Cores deliver substantial performance gains for stencils, contradicting the long-standing consensus that they are not suitable for these memory-bound tasks.

The aforementioned paradox raises a critical research question: 
\textbf{Why Tensor Cores benefit stencil computation, a problem conventionally recognized as memory-bound? And what is the underlying mechanism defying the memory wall?}

To answer these questions, this paper presents an in-depth performance analysis that demystifies the acceleration mechanism of Tensor Cores for stencil computations. 
By revisiting the interplay between stencil characteristics and Tensor Core constraints, we characterize the algorithmic inefficiencies and establish an enhanced roofline model to estimate the stencil performance on different hardware units. This model guides a systematic analysis to delineate the \textit{sweet spot} where Tensor Cores outperform general-purpose units.
Ultimately, our analysis resolves the initial contradiction by demonstrating that Tensor Cores can indeed \textit{break the performance ceiling} of conventional approaches in certain scenarios.

In summary, this paper makes the following contributions:

\begin{itemize}
    \item We dissect the architectural mismatches between stencil patterns and Tensor Core constraints, specifically analyzing the performance impact of using Tensor Cores, including overheads from operand alignment and kernel fusion.

    \item An enhanced performance model that integrates these overheads is proposed to formulate stencil workloads, enabling a fair comparison of effective throughput between Tensor Cores and conventional CUDA Cores.

    \item We derive analytical criteria to identify the \textit{sweet spot} where Tensor Core acceleration is theoretically profitable, and further demonstrate how Sparse Tensor Cores expand this effective design space.

    \item We validate our model through extensive evaluations across SOTA implementations, including DRStencil, EBISU, ConvStencil, and SPIDER, validating the effectiveness of our performance model and analytical criteria.
\end{itemize}

\begin{table}[b]\small
    \centering
    \rowcolors{2}{gray!10}{white}
    \begin{tabular}{c|c}
    \toprule
    \textbf{Symbol} & \textbf{Description} \\
    \midrule
    $d$             & Dimensionality of Stencil Pattern. \\
    $r$             & Radius (or Order) of Stencil Pattern.\\
    $\mathbb{P}$    & Peak Performance of Specific Hardware.\\
    $\mathbb{B}$    & Memory Bandwidth of Specific Hardware.\\
    $t$             & Fusion Depth. \\
    ${C}$           & Computational Workload.\\
    ${M}$           & Memory Traffic.\\
    ${I}$           & Arithmetic Intensity.\\
    ${K}$           & Number of Points in the Stencil Kernel. \\
    ${P}$           & Attainable Performance. \\
    $\mathbb{S}$    & Sparsity Introduced by Algorithmic Transformation. \\
    $\alpha$        & Redundancy Factor Introduced by Kernel Fusion.\\
    \bottomrule
    \end{tabular}
    \caption{Nomenclature.}
    \label{tab.notation}
\end{table}

\section{What Happened When Introducing Tensor Cores?}

To resolve the aforementioned contradiction, we must first understand how stencil computations are adapted onto Tensor Cores. Since Tensor Cores are designed for matrix multiplication, a workload distinct from stencil computation, adapting stencils onto these units is non-trivial. It requires transformations to satisfy constraints regarding tensor contraction and operand size. These transformations, while enabling execution, often introduce additional overheads. This section clarifies these hardware constraints, the corresponding transformations, and the associated overheads.

\subsection{Constraints Imposed by Tensor Cores}

\subsubsection{Tensor Contraction Constraint.}

Tensor Cores are specialized 2D ALUs introduced by NVIDIA to accelerate matrix multiply-and-accumulate (MMA) operations~\cite{amperewhitepaper}. Similar ALUs have been adopted by other vendors, including AMD's Matrix Cores~\cite{amdmatrixcore} and Intel's Advanced Matrix Extensions (AMX)~\cite{intelamx}. Despite minor architectural differences, these units implement the same tensor contraction: $\mathbb{R}^{m\times k} \times \mathbb{R}^{k\times n}\rightarrow \mathbb{R}^{m\times n}$, which can be formalized as:
\begin{equation}
    D_{m \times n} = A_{m \times k} \times B_{k \times n} + C_{m \times n},
\end{equation}
This operation is often denoted as $\texttt{m}{M}\texttt{n}{N}\texttt{k}{K}$, where each dimension is followed by its size parameter. For example, \texttt{m16n8k16} represents an MMA operation with dimensions $m=16, n=8, k=16$.

\subsubsection{Operand Size Constraint.}\label{sec:operand_size_constraint}

Unlike conventional ALUs, such as CUDA Cores, which operate on calculations with fine-grained threads, Tensor Cores operate at a coarser granularity. They process fixed-size matrix tiles collectively across a group of threads (e.g., warp). 
Specifically, modern GPUs define fundamental MMA instruction shapes, the minimum atomic units of computation, including \texttt{m8n8k4} for double, \texttt{m16n8k16} for \texttt{float}~\cite{nvidiaisa}. Matrices smaller than these primitive dimensions must be padded or restructured to conform to hardware requirements, often incurring non-negligible overheads in both memory footprint and computational efficiency.

\subsection{Adapting Stencil onto Tensor Cores} \label{sec.redundant_computation}

\subsubsection{Bridging Tensor Contraction Mismatch.}

\begin{figure}[t]
  \includegraphics[width=\linewidth]{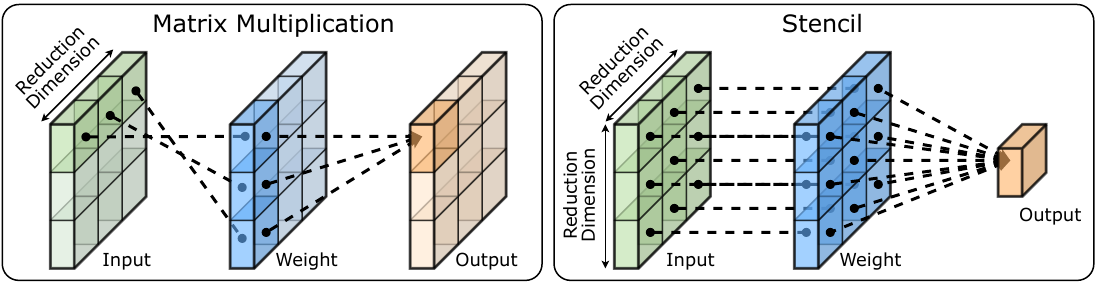} 
  \caption{Reduction Dimension Mismatch between Matrix Multiplication and Stencil.}
  \label{fig.reduction_dimension_mismatch}
\end{figure}

As described above, Tensor Cores perform matrix multiplication by contracting tensors along a single shared reduction dimension. In contrast, stencil computations involve localized contractions along multiple dimensions, determined by the dimensionality $d$. Figure \ref{fig.reduction_dimension_mismatch} shows this structural mismatch between matrix multiplication and Box2D stencil.

\begin{figure}[t]
\centering
\begin{subfigure}[b]{\linewidth}
    \centering
    \includegraphics[width=\linewidth]{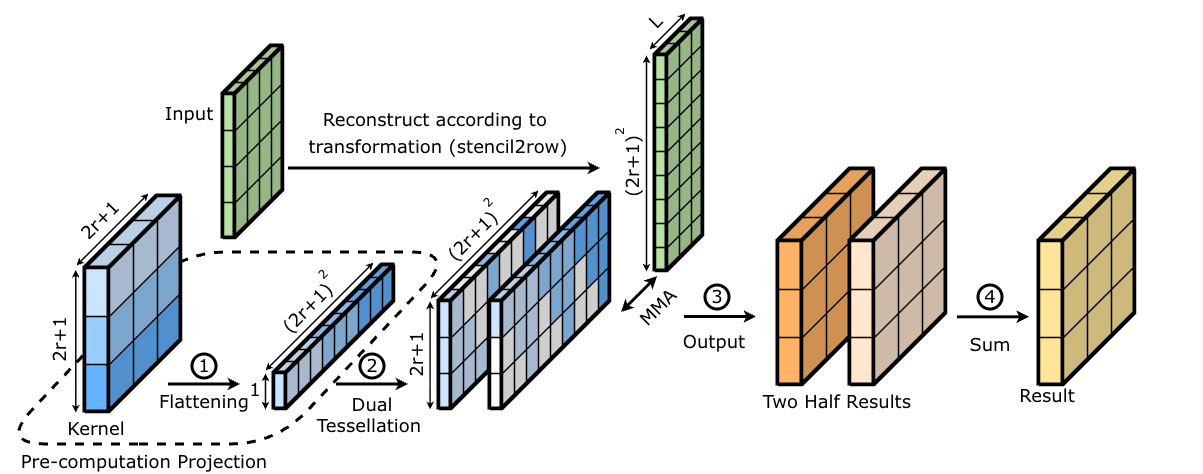}
    \caption{Flattening-based Transformation Scheme}
    \label{fig:flatten_scheme}
\end{subfigure}

\begin{subfigure}[b]{\linewidth}
    \centering
    \includegraphics[width=\linewidth]{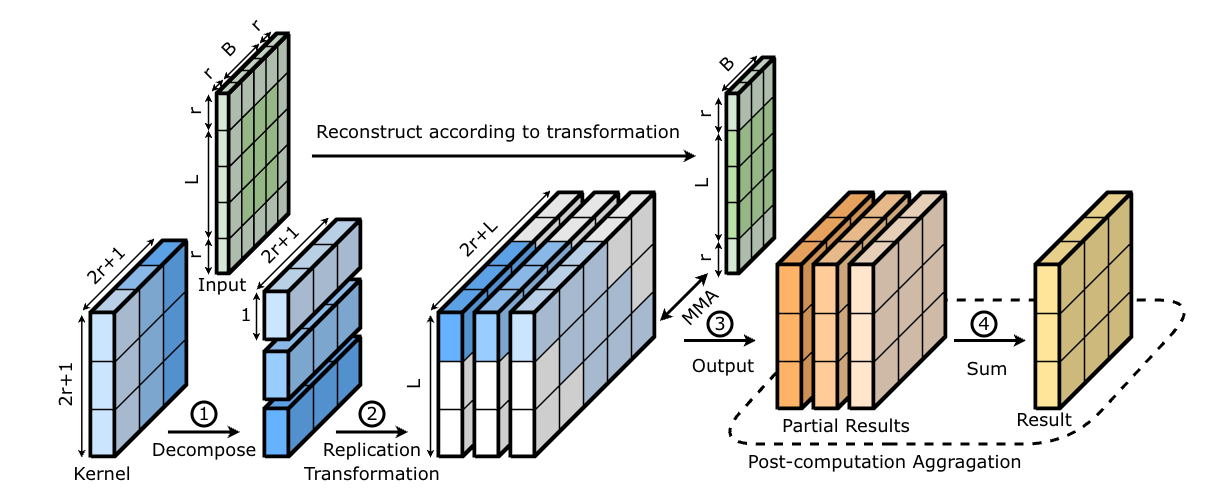}
    \caption{Decomposing-based Transformation Scheme}
    \label{fig:decompose_scheme}
\end{subfigure}

\caption{Two Typical Transformation Schemes to Adapt Stencil Computation onto Tensor Cores.}
\label{fig:Transform_to_tensor_core}
\end{figure}

To reconcile this mismatch, prior works have proposed various transformation strategies, which generally fall into two categories:

(1) \textbf{Flattening}: As shown in the step \ding{192} of Figure \ref{fig:flatten_scheme}, this approach transforms the kernel via flattening. The multi-dimensional stencil weights are linearized along the MMA reduction dimension (similar to \texttt{img2col} algorithm). By performing this projection before invoking Tensor Cores, it maps all spatial reduction dimensions onto the single GEMM reduction axis, enabling the Tensor Core to perform the contraction atomically. ConvStencil~\cite{chen2024convstencil} is a representative work employing this approach.

(2) \textbf{Decomposing}: As illustrated in the step \ding{192} of Figure \ref{fig:decompose_scheme}, this approach transforms the kernel via decomposition. The stencil kernel is split into multiple independent vectors aligned with the reduction dimensions of Tensor Cores. The partial results generated from each vector are subsequently accumulated in Step \ding{195}. This post-computation aggregation enables Tensor Cores to complete the full stencil contraction. Representative works employing this approach include TCStencil~\cite{TCStencil}, LoRAStencil~\cite{zhang2024lorastencil} and SPIDER~\cite{SPIDER}.

\subsubsection{Aligning with Operand Size Requirement.}\label{sec:sparse_in_transformation}

Satisfying the contraction constraint is necessary but insufficient. While the strategies in Step \ding{192} successfully transform the computation to MMA form, they typically yield vectors with a dimension of $m = 1$.

This creates a severe conflict with the hardware's operand size constraint. As noted in \S \ref{sec:operand_size_constraint}, Tensor Cores impose a minimum height requirement (e.g., $m=8$ for \texttt{m8n8k4}), directly using these vectors results in extreme under-utilization (e.g., utilizing only $1/8 = 12.5\%$ of compute resource).

To mitigate this, recent methods all employ an additional transformation, such as \textit{dual tessellation} in ConvStencil (step \ding{193} in Figure \ref{fig:flatten_scheme}) and \textit{replication} in SPIDER (step \ding{193} in Figure \ref{fig:decompose_scheme}). These transformations expand single vectors into matrices that satisfy the minimum operand size requirement.

\begin{figure}[t]
  \includegraphics[width=0.88\linewidth]{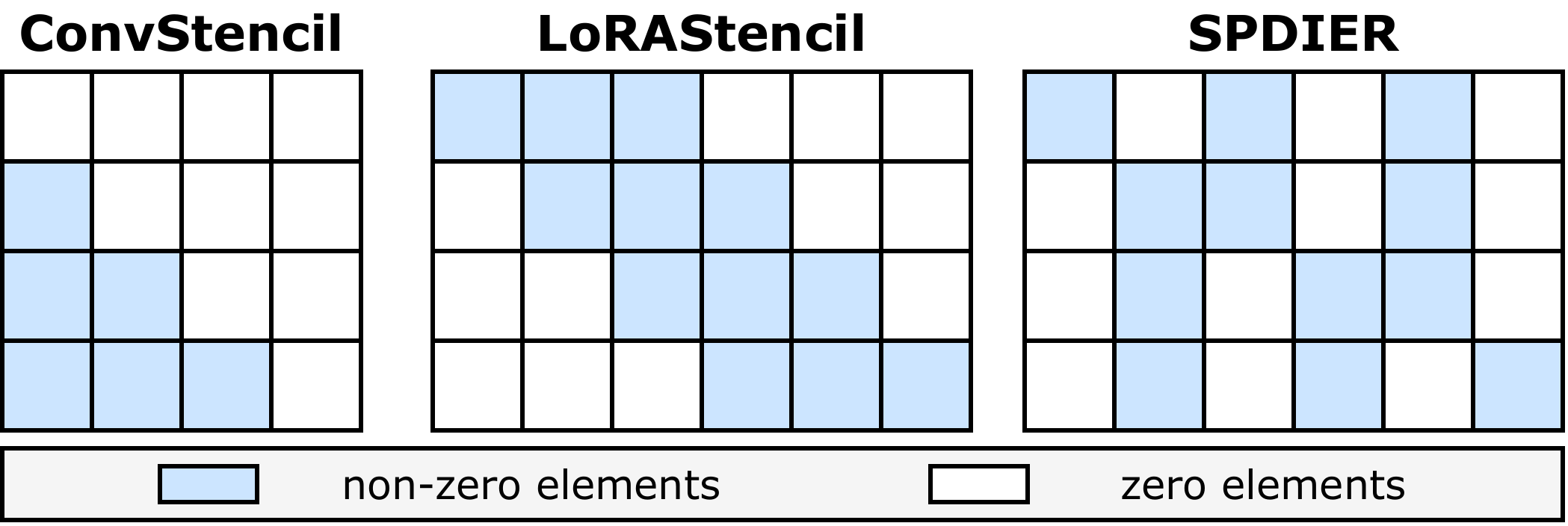} 
  \caption{Transformed Sparse Matrices in Recent Tensor Core-based  Implementations.}
  \label{fig.sparse_matrix}
\end{figure}

However, this transformation comes at a cost. As illustrated in Figure \ref{fig.sparse_matrix}, constructing these matrices from the original stencil kernel inevitably requires padding with zeros to maintain mathematical equivalence. These padded values introduce redundant operations that do not contribute to the actual computation. We characterize this phenomenon as \textit{sparse redundancy}, where the hardware operates on sparse matrices, thus wasting compute resource.

\textbf{Notation.} 
To quantify this overhead, let $C$ denote the number of actual operations for stencil computation (excluding padding), and $\mathbb{S} \in (0, 1]$ denote the \emph{sparsity factor} of the transformed matrix (fraction of non-zero elements). The number of operations executed on Tensor Cores $C_{TC}$ ($TC$ is short for Tensor Core) is:
\begin{equation}
    {C}_{TC} = \frac{C}{\mathbb{S}}
\end{equation}
For example, if 50\% of the transformed matrix is zero, the executed operations are twice the ideal workload.

\subsubsection{Handling Stencils with Small Radius.}

The aforementioned adaptations enable Tensor Cores to compute stencil problems efficiently for most cases. However, this efficiency degrades significantly for small-radius stencils. To be specific, for the flattening scheme, a small radius $r=1$ produces a transformed matrix of dimensions $m=3,n=9$. To satisfy the operand size constraint (e.g., $m \ge 8$), the dimension $m$ must be padded, wasting $62.5\%$ of compute resource.
The decomposition scheme suffers from similar limitations, as the sparsity factor $\mathbb{S}$ increases with the growth of stencil radius $r$. This implies that a small radius yields excessively sparse matrices. For example, with $r=1$, about $62.5\%$ of matrix entries are zero-padded values.

To mitigate this inefficiency, recent works adopt a kernel fusion technique to aggregate multiple time steps into a single monolithic step, thereby increasing the stencil kernel size. While conceptually similar to temporal fusion, the underlying execution mechanisms diverge fundamentally.

Specifically, conventional temporal fusion on CUDA Cores processes time steps sequentially, allowing for the explicit reuse of intermediate results. In contrast, Tensor Core-based kernel fusion encapsulates all time steps into a single monolithic matrix operation. This design precludes the reuse of intermediate values, inevitably leading to redundant arithmetic operations. Consequently, while this strategy resolves the operand size mismatch, it introduces a new source of overhead, which we characterize as \textit{fusion redundancy}.

\begin{figure}[t]
  \includegraphics[width=0.88\linewidth]{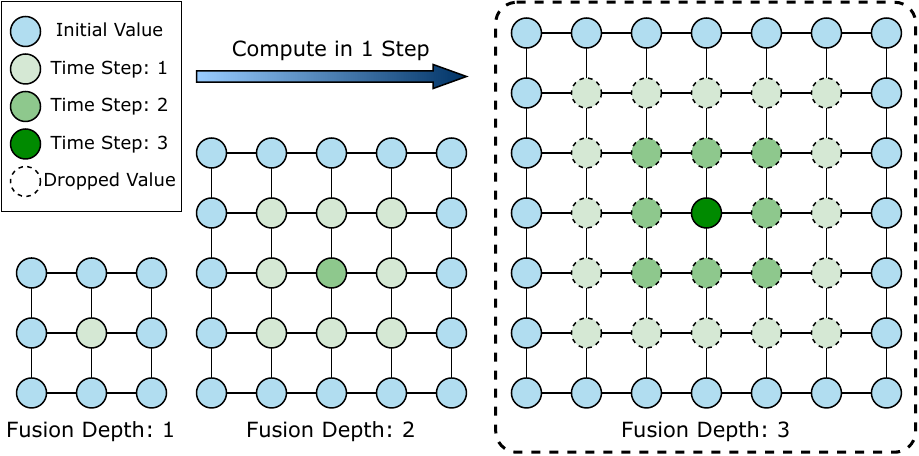} 
  \caption{Illustration of Kernel Fusion Strategy for Stencil Computations on Tensor Cores.}
  \label{fig.fusion_redundant}
\end{figure}

Figure~\ref{fig.fusion_redundant} illustrates this effect for a \textbf{Box-2D1R} stencil with a fusion depth of $3$. The fused kernel operates on a $7 \times 7$ spatial points, requiring $49$ multiply-and-add operations to produce the central output in a single step. In contrast, a sequentially executed implementation would conduct $3\times3$ operations updating each step, requiring only $27$ operations in total for $3$ updates. Thus, kernel fusion introduces $22$ redundant operations per output point.

\textbf{Notation.} We characterize this overhead by introducing a \emph{fusion redundancy factor} $\alpha$ (formally derived in \S\ref{sec.tc_formulation}), which captures the increase in compute operations due to kernel fusion. The computational workload on Tensor Cores with $t$ fusion steps becomes:

\begin{equation} \label{eq.compute_ops}
    {C}_{TC}^{(t)} = \frac{\alpha}{\mathbb{S}}\cdot C^{(t)}
\end{equation}

\section{Performance Formulation}

In this section, we formulate an analytical performance model for stencil computations on both CUDA Cores and Tensor Cores, aiming to characterize their distinct execution behaviors across different hardware.

\subsection{Roofline Model}

The roofline model serves as a canonical framework for projecting the upper-bound performance of a specific workload~\cite{rooflinemodel}. It characterizes the attainable performance ${P}$ (in FLOPS) as a function of arithmetic intensity ${I}$ (in FLOPs/Byte), which represents the ratio of computational operations ${C}$ (in FLOPs) to data traffic ${M}$ (in Bytes):
\begin{equation}
    {I} = \frac{{C}}{{M}}
\end{equation}

\begin{figure}[t]
  \includegraphics[width=0.88\linewidth]{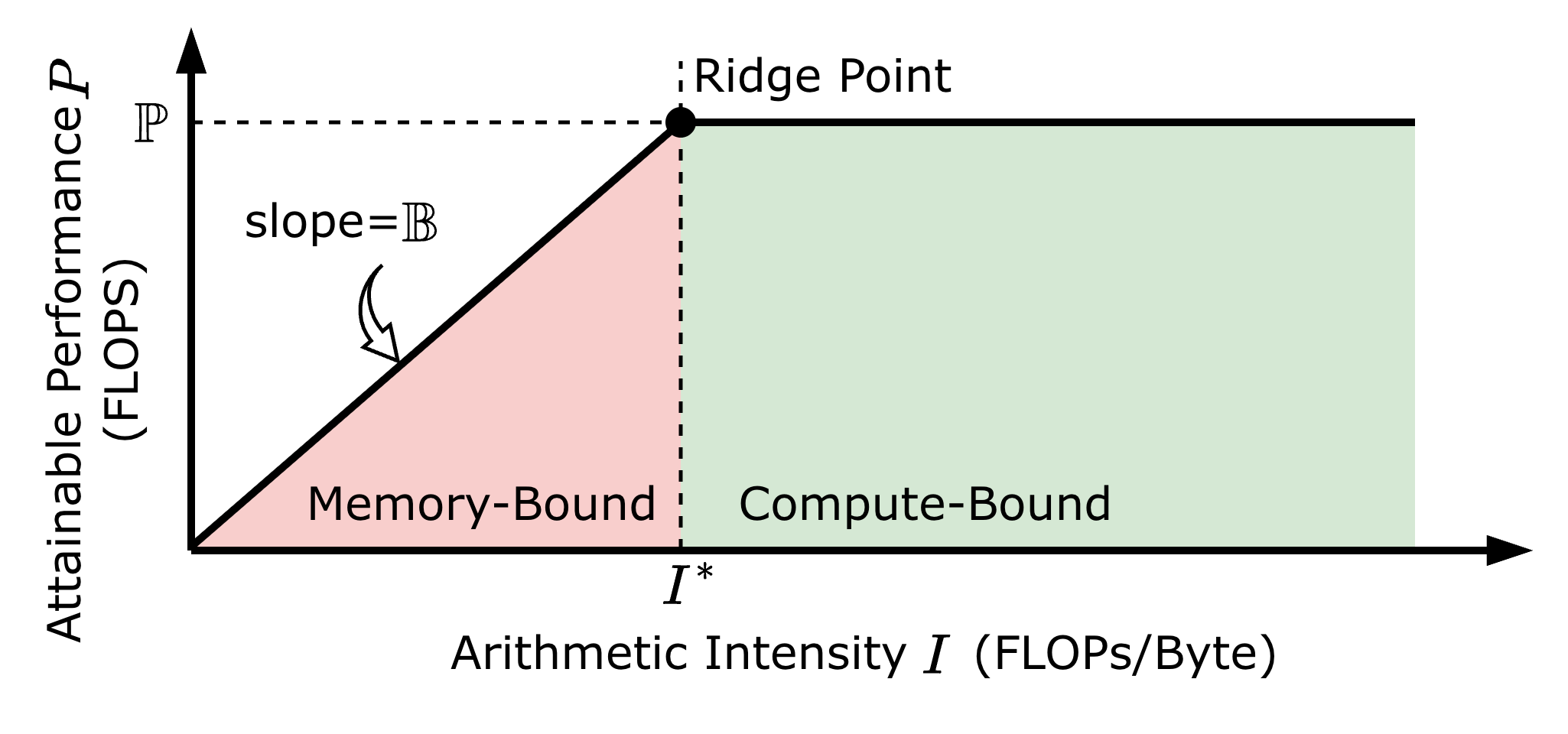} 
  \caption{Roofline Model Illustration.}
  \label{fig.roofline_model}
\end{figure}

As illustrated in Figure \ref{fig.roofline_model}, the model establishes a performance roofline comprising two distinct regions, delineated by the ridge point $I^*$. In the memory-bound region (${I} < I^*$), performance scales linearly with arithmetic intensity, with the slope determined by the memory bandwidth $\mathbb{B}$ of the hardware (in Bytes/sec). Conversely, in the compute-bound region (${I} > I^*$), performance is throttled by the peak compute throughput $\mathbb{P}$ of the hardware (in FLOPS). Thus, the attainable performance $P$ is formulated as:
\begin{equation}
    {P} = \min(\mathbb{P},\; \mathbb{B} \cdot {I})
\end{equation}

This framework provides a rigorous metric to diagnose performance bottlenecks, enabling us to identify whether an implementation is limited by memory bandwidth or compute resource.

\subsection{Performance on Different Hardware}\label{sec.formulation}

\subsubsection{Original Stencil Problem}

We begin by modeling the performance of the original stencil problem. Let ${K}$ denote the number of spatial points in the stencil pattern. For example, a box-shaped stencil with dimensionality $d$ and radius $r$ should have ${K}=(2r+1)^d$ neighboring points. 

Each output point is updated via a fused multiply-add (FMA) operation per neighboring point, which corresponds to two floating-point operations (one multiply and one add). Thus, the total number of floating-point operations required to update a single point can be expressed as ${C} = 2{K}$.

Regarding memory traffic, let ${D}$ represent the size (in Bytes) of the data type (e.g., ${D} = 4$ for \texttt{float}). In an ideal implementation, each update reads one input value and writes one output value, resulting in a total memory traffic of ${M} = 2{D}$ bytes per point.

The arithmetic intensity ${I}$ of this problem is therefore:
\begin{equation}
    {I} = \frac{{C}}{{M}} = \frac{{K}}{{D}}
\end{equation}

Finally, the attainable performance ${P}$ is bounded by both the peak compute throughput $\mathbb{P}$ and the memory bandwidth $\mathbb{B}$, yielding:  
\begin{equation}
    {P} = \min(\mathbb{P}, \mathbb{B} \cdot {I}), \ {I} = \frac{{C}}{{M}} = \frac{{K}}{{D}}
\end{equation}

It should be noted that, when implementing stencil computations on GPUs, halo regions may overlap across adjacent tiling blocks. For conciseness, we omit the additional memory accesses and redundant computations incurred by these overlapping elements in our model.

\subsubsection{CUDA Core Implementation with Temporal Fusion}

The CUDA Core implementation follows the original problem formulation, with computational cost ${C}_{CU} = {C}$ and memory traffic ${M}_{CU} = {M}$ per output point, where $CU$ is short for CUDA Cores. 

To improve performance, temporal fusion is commonly employed to enhance data locality. By fusing $t$ time steps into a single step, the computational workload is increased by a factor of $t$, yielding ${C}_{CU}^{(t)} = t{C}$. Crucially, the intermediate result from earlier steps can be retained in fast on-chip memory (e.g., shared memory or registers), thereby avoiding additional off-chip memory traffic. As a result, the memory traffic remains ${M}_{CU}^{(t)} = {M}$.

Consequently, the arithmetic intensity and attainable performance for temporally fused CUDA Core implementation with fusion depth $t$ are given by:

\begin{equation}\label{eq.cuda_t_ai}
\begin{aligned}
    {I}^{(t)}_{CU} &= \frac{{C}^{(t)}_{CU}}{{M}^{(t)}_{CU}} = t\cdot\frac{{K}}{{D}} \\
    {P}_{CU}^{(t)} &= \min(\mathbb{P}_{CU}, \mathbb{B}\cdot {I}^{(t)}_{CU})
\end{aligned}
\end{equation}
where $\mathbb{P}_{CU}$ denotes the peak compute throughput of the CUDA Core hardware.

\subsubsection{Tensor Core Implementation with Kernel Fusion} \label{sec.tc_formulation}

When adapting stencil computation onto Tensor Cores, the memory traffic per output point remains unchanged, with ${M}_{{TC}}^{(t)} = {M}$. However, the computational workload is significantly affected by two key factors introduced during adaptation: sparsity factor $\mathbb{S}$ and redundancy factor $\alpha$, as characterized in Equation~\ref{eq.compute_ops}.

Here, the sparsity factor $\mathbb{S}$ arises from hardware-imposed operand size constraints. Being a transformation-specific constant, $\mathbb{S}$ lacks a universal formulation, instead, it must be derived individually based on the underlying transformation scheme.

Meanwhile, the redundancy factor $\alpha$ stems from kernel fusion. It quantifies the inflation in the number of spatial points due to fusing $t$ time steps into a single monolithic kernel, which can be derived as: 
\begin{equation}\label{eq.alpha}
    \alpha = \frac{{K}^{(t)}}{t\cdot{K}}
\end{equation} 
where ${K}^{(t)}$ denotes the number of spatial points in the enlarged stencil kernel after fusion. Fusing $t$ time steps expands the effective stencil radius from $r$ to $t \cdot r$, and the exact form of $\alpha$ depends on the stencil shape.

As an example, consider a box-shaped stencil in $d$ dimensions, the number of kernel points is ${K} = (2r+1)^d$, while the fused kernel spans ${K}^{(t)} = (2r \cdot t+1)^d$. Thus, the redundancy factor becomes:
\begin{equation}\label{eq.alpha_box}
    \alpha_{box} = \frac{{K}^{(t)}}{t{K}} = \frac{(2r \cdot t+1)^d}{t \cdot (2r+1)^d}
\end{equation}

Let $\mathbb{P}_{TC}$ denote the peak compute throughput of the Tensor Core hardware. The arithmetic intensity ${I}^{(t)}_{TC}$ and raw performance of the fused Tensor Core implementation ${P}_{TC}^{(t)}$ are given by:
\begin{equation}\label{eq.tc_t_ai_1}
\begin{aligned}
    {I}^{(t)}_{TC} &= \frac{{C}^{(t)}_{TC}}{{M}^{(t)}_{TC}} = t\cdot\frac{\alpha}{\mathbb{S}}\cdot\frac{{K}}{{D}} \\
    {P}_{TC}^{(t)} &= \min(\mathbb{P}_{TC}, \mathbb{B}\cdot {I}^{(t)}_{TC})
\end{aligned}
\end{equation}

Importantly, the performance ${P}_{TC}^{(t)}$ takes every operation into account, including the redundant ones (computing on padded values). Therefore, we must normalize this performance to get the actual performance ${P}_{TC,actual}^{(t)}$, yielding:
\begin{equation}\label{eq.tc_t_ai}
\begin{aligned}
    {I}^{(t)}_{TC} &= \frac{{C}^{(t)}_{TC}}{{M}^{(t)}_{TC}} = t\cdot\frac{\alpha}{\mathbb{S}}\cdot\frac{{K}}{{D}} \\
    {P}_{TC,actual}^{(t)} &= \frac{{P}_{TC}^{(t)}}{\frac{\alpha}{\mathbb{S}}} = \frac{\mathbb{S}}{\alpha} \cdot \min(\mathbb{P}_{TC}, \mathbb{B}\cdot {I}^{(t)}_{TC})
\end{aligned}
\end{equation}

\section{How Can Tensor Cores Accelerate Stencil?}

\subsection{Performance Comparative Analysis} \label{sec.analysis}

\begin{figure*}[t]
    \centering
    \begin{subfigure}[b]{0.26\textwidth}
        \adjustbox{valign=b, height=3.2cm}{\includegraphics{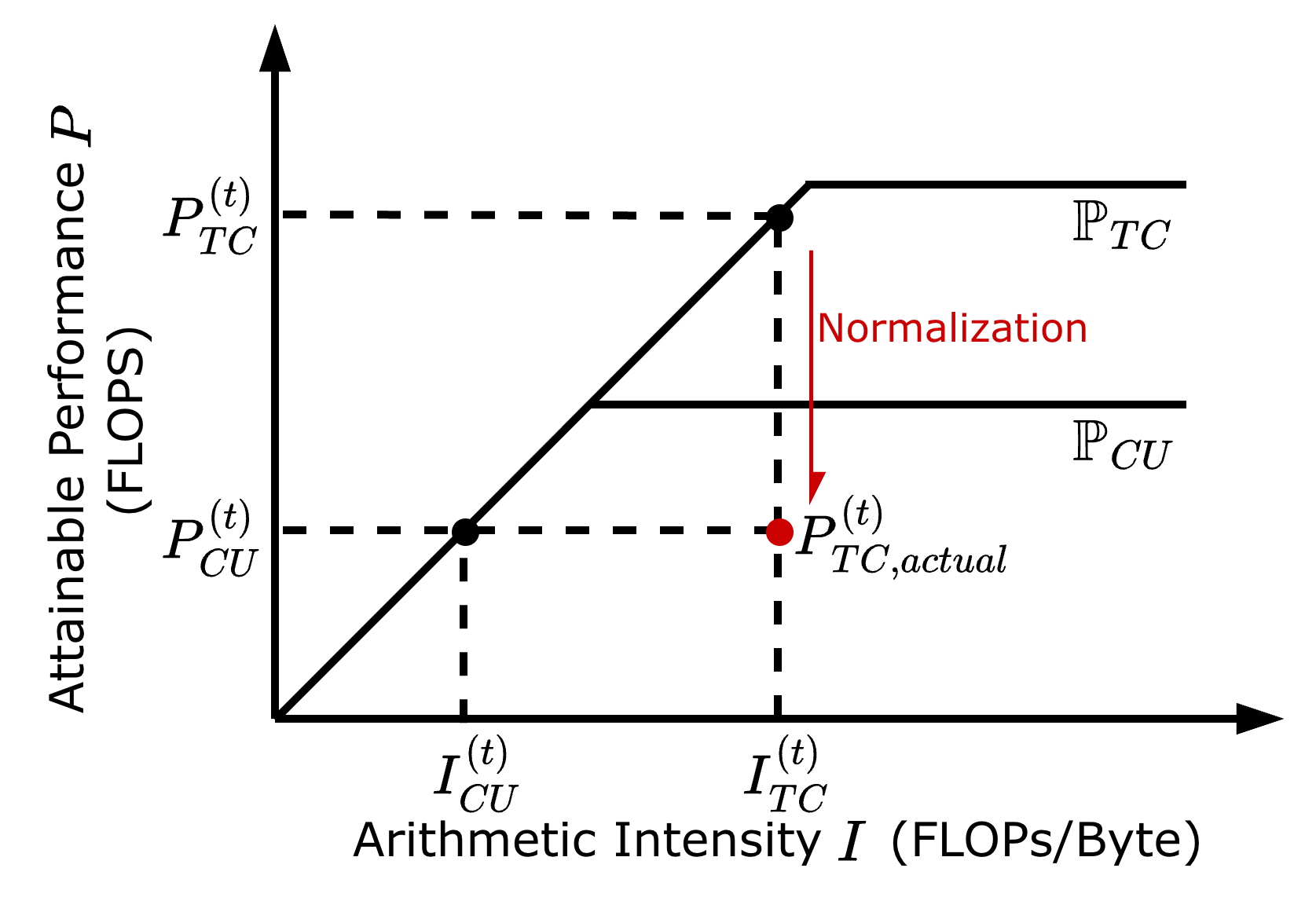}}
        \caption{CU: MB $\rightarrow$ TC: MB}
        \label{fig.case1}
    \end{subfigure}
    \begin{subfigure}[b]{0.235\textwidth}
        \adjustbox{valign=b, height=3.2cm}{\includegraphics{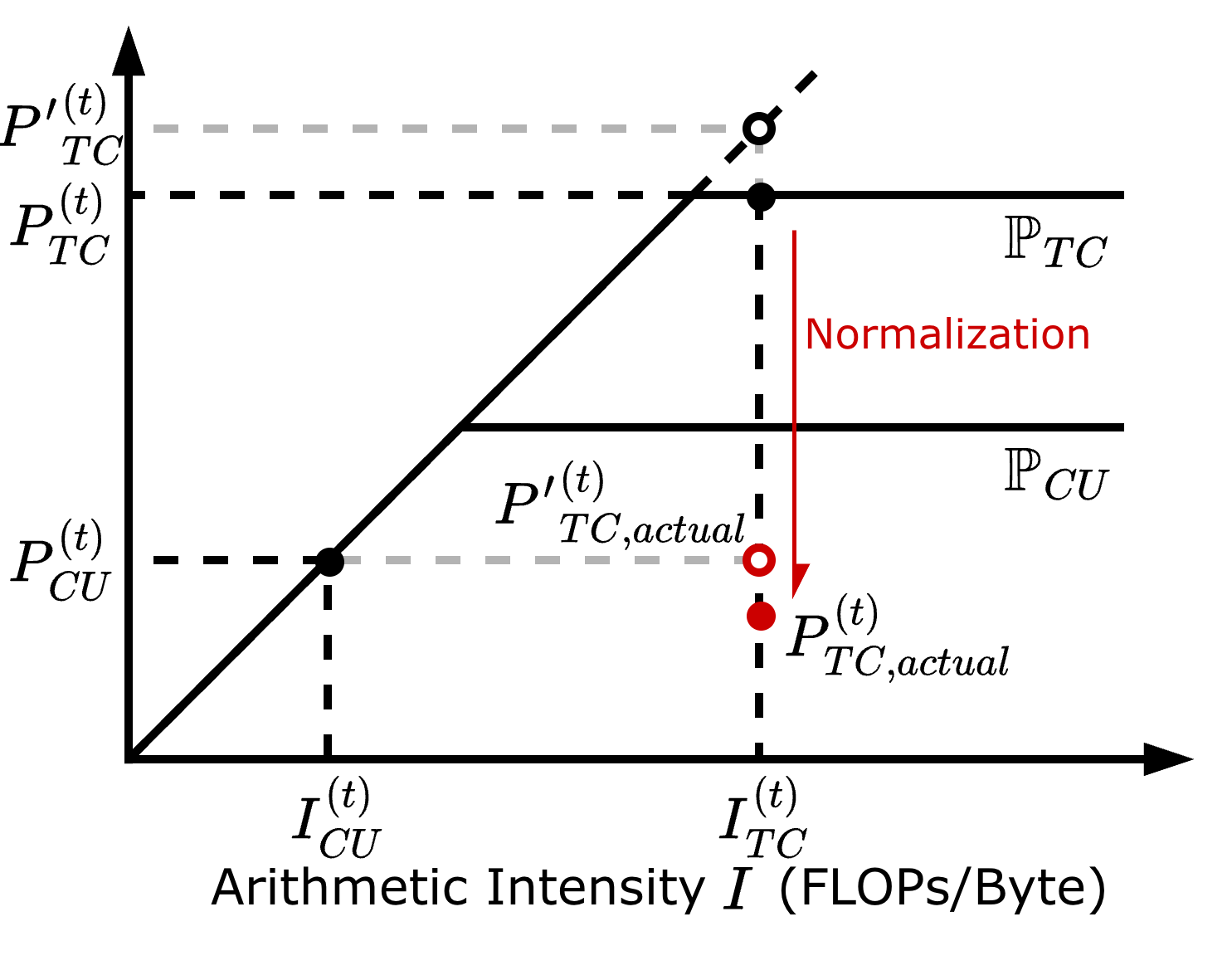}}
        \caption{CU: MB $\rightarrow$ TC: CB}
        \label{fig.case2}
    \end{subfigure}
    \begin{subfigure}[b]{0.235\textwidth}
        \adjustbox{valign=b, height=3.2cm}{\includegraphics{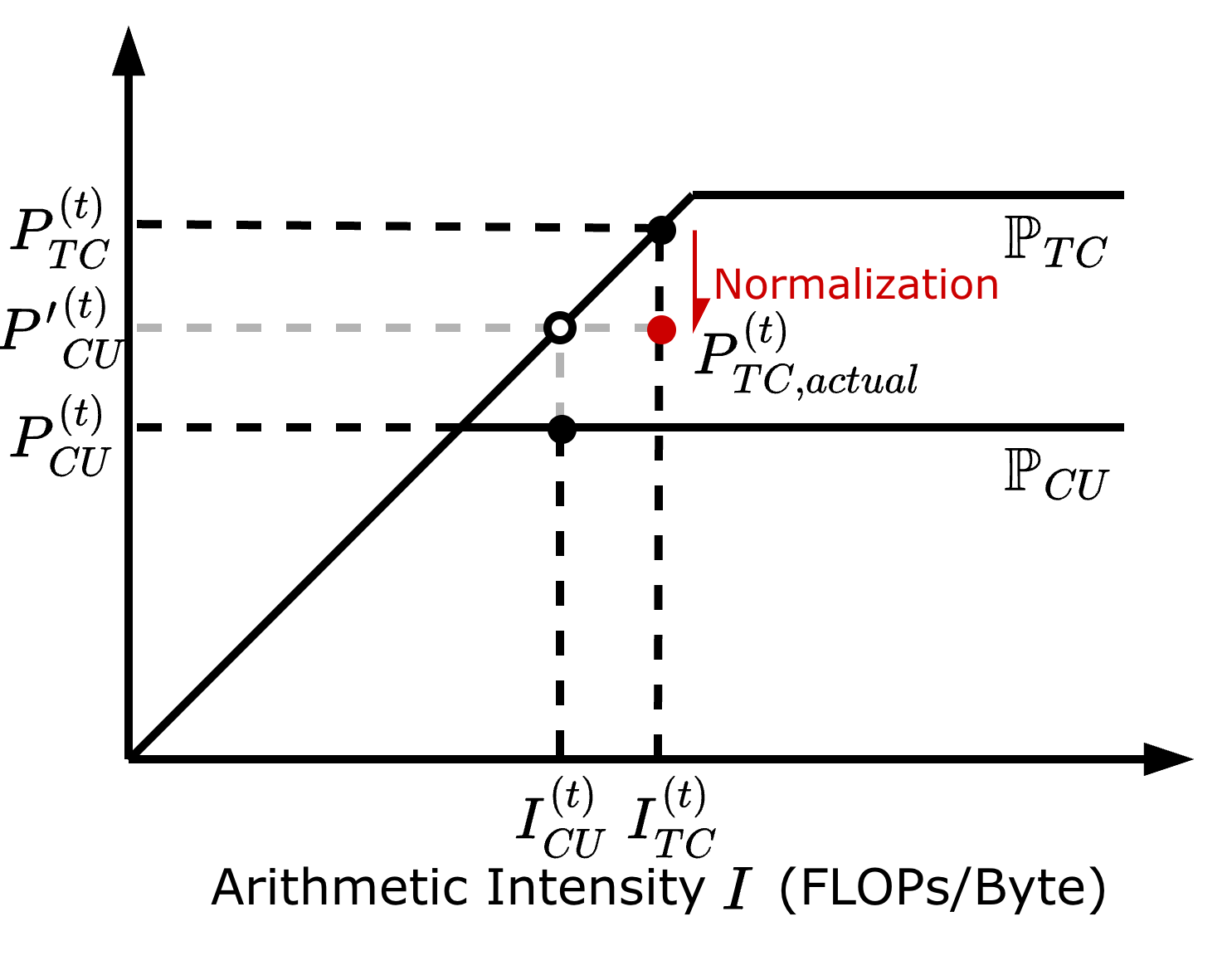}}
        \caption{CU: CB $\rightarrow$ TC: MB}
        \label{fig.case3}
    \end{subfigure}
    \begin{subfigure}[b]{0.235\textwidth}
        \adjustbox{valign=b, height=3.2cm}{\includegraphics{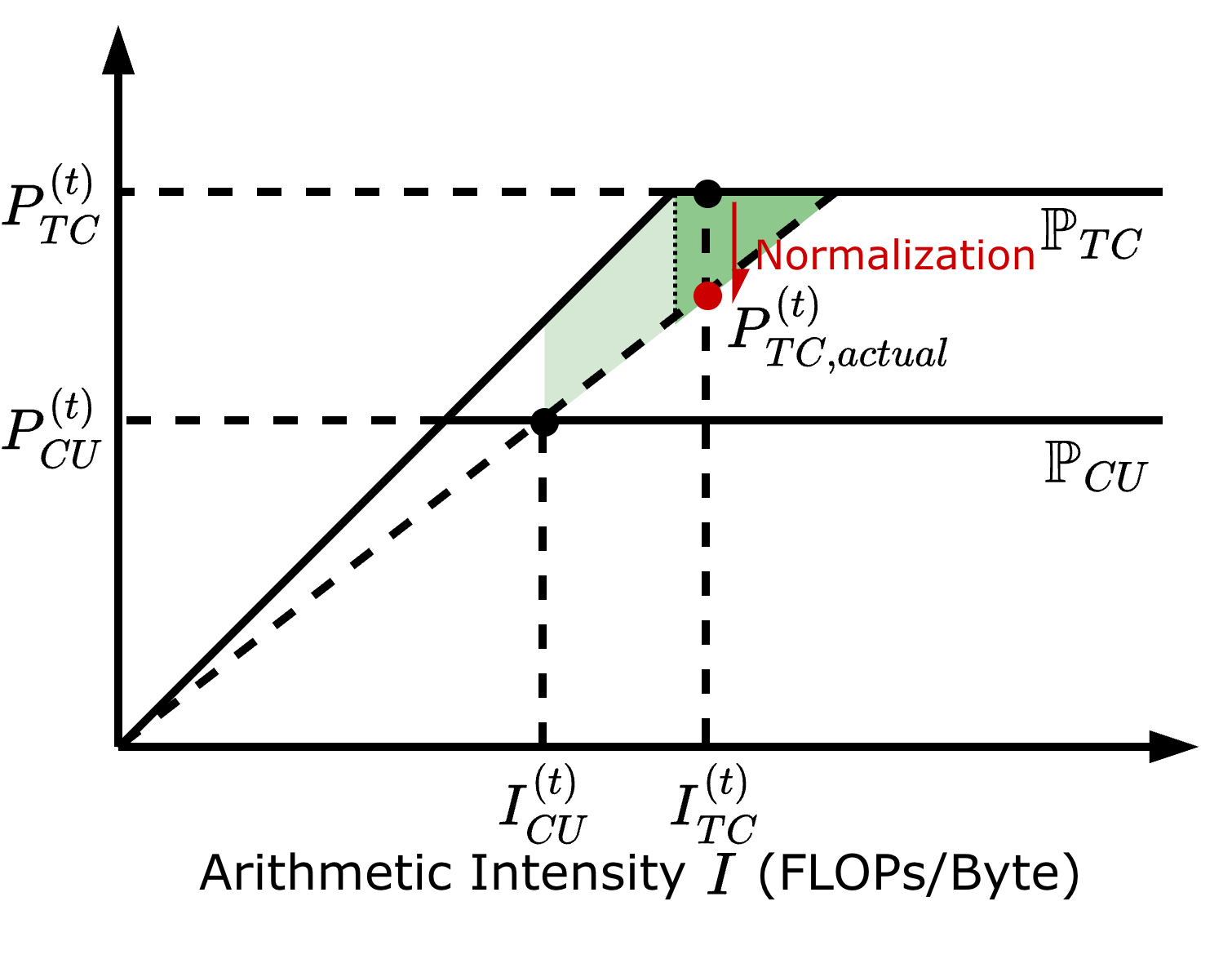}}
        \caption{CU: CB $\rightarrow$ TC: CB}
        \label{fig.case4}
    \end{subfigure}
    \caption{Illustration of Stencil Performance on CUDA Cores and Tensor Cores under Four Scenarios. $MB$ denotes memory-bound problems and $CB$ represents compute-bound problems.}
    \label{fig.example}
\end{figure*}

With the performance model established, we assess the acceleration potential of Tensor Cores for stencil computations by comparing their effective performance $P_{TC,actual}^{(t)}$ against the baseline CUDA Cores $P_{CU,actual}^{(t)}$.
Since CUDA Cores execution incurs no computational redundancy, its effective performance $P_{CU,actual}^{(t)}$ aligns directly with its theoretical throughput $P_{CU}^{(t)}$. Leveraging Equation \ref{eq.cuda_t_ai} and Equation \ref{eq.tc_t_ai}, we formulate the speedup ratio as:

\begin{equation}
\begin{aligned}
        \frac{P_{TC,actual}^{(t)}}{P_{CU,actual}^{(t)}} 
        & = \frac{\frac{\mathbb{S}}{\alpha} \cdot \min(\mathbb{P}_{TC}, \mathbb{B}\cdot {I}^{(t)}_{TC})}{\min(\mathbb{P}_{CU}, \mathbb{B}\cdot {I}^{(t)}_{CU})} \\
        & = \frac{\frac{\mathbb{S}}{\alpha} \cdot \min(\mathbb{P}_{TC}, \mathbb{B}\cdot t \cdot \frac{\alpha}{\mathbb{S}} \cdot \frac{{K}}{{D}})}{\min(\mathbb{P}_{CU}, \mathbb{B}\cdot t \cdot \frac{{K}}{{D}})}
\end{aligned}
\end{equation}

This ratio reflects the speedup of adapting stencils onto Tensor Cores compared to conventional CUDA Core implementations. A ratio exceeding 1 indicates that Tensor Cores deliver effective acceleration.

Given the piecewise nature of the underlying performance models, we linearize the performance model to rigorously evaluate it by partitioning our analysis into four orthogonal scenarios. These scenarios are strictly defined by the memory-bound and compute-bound states of the comparison units.

\textbf{(1) Memory-bound on CUDA Cores $\rightarrow$ Memory-bound on Tensor Cores.} 

When both implementations operate within the memory-bound region, the peak throughput is strictly throttled by the memory bandwidth $\mathbb{B}$. The performance ratio simplifies to:

\begin{equation}
\frac{P_{TC,actual}^{(t)}}{P_{CU,actual}^{(t)}} = \frac{\frac{\mathbb{S}}{\alpha} \cdot \mathbb{B}\cdot t \cdot \frac{\alpha}{\mathbb{S}} \cdot \frac{{K}}{{D}}}{\mathbb{B}\cdot t \cdot \frac{{K}}{{D}}} = 1
\end{equation}

This indicates that the effective stencil performance is \textbf{ equivalent} on both CUDA Cores and Tensor Cores under the memory-bandwidth constraints. As illustrated in Figure~\ref{fig.case1}, the superior throughput provided by Tensor Cores is entirely offset by the overhead of processing redundant zero-padded elements.

\textbf{(2) Memory-bound on CUDA Cores $\rightarrow$ Compute-bound on Tensor Cores.}

In this scenario, the CUDA Core baseline is memory-bound. Conversely, the transformations required for Tensor Cores inflate the arithmetic intensity, driven by sparsity and redundancy factors, thereby shifting the execution into the compute-bound region. Consequently, the speedup ratio is derived as:

\begin{equation}
\frac{P_{TC,actual}^{(t)}}{P_{CU,actual}^{(t)}} = \frac{\frac{\mathbb{S}}{\alpha} \cdot \mathbb{P}_{TC}}{B\cdot t \cdot \frac{{K}}{{D}}}
\end{equation}

Under this scenario ($\mathbb{P}_{TC} < \mathbb{B}\cdot I_{TC}^{(t)}$), the Tensor Core implementation is limited by compute resource, implying that:

\begin{equation} \label{eq.MB_to_CB_compare}
\frac{P_{TC,actual}^{(t)}}{P_{CU,actual}^{(t)}} = \frac{\frac{\mathbb{S}}{\alpha} \cdot \mathbb{B}\cdot I_{TC}^{(t)}}{\mathbb{B} \cdot t \cdot \frac{{K}}{{D}}} < \frac{\frac{\mathbb{S}}{\alpha} \cdot \mathbb{B}\cdot t \cdot \frac{\alpha}{\mathbb{S}} \cdot \frac{{K}}{{D}}}{\mathbb{B} \cdot t \cdot \frac{{K}}{{D}}} = 1
\end{equation}

As a result, Tensor Core implementation consistently \textbf{underperforms} CUDA Core implementation in this scenario. As illustrated in Figure~\ref{fig.case2}, achieving equivalent stencil performance on CUDA Cores would necessitate a theoretical throughput of ${P'}_{TC}^{(t)}$ on Tensor Cores, a value that exceeds the physical hardware ceiling. Consequently, the effective performance on Tensor Cores is penalized by this computational limit, failing to match the performance on CUDA Cores.

\textbf{(3) Compute-bound on CUDA Cores $\rightarrow$ Memory-bound on Tensor Cores.}

Although adapting stencils to Tensor Cores increases arithmetic intensity, the workload may paradoxically shift from compute-bound to memory-bound. This occurs because the higher peak throughput of Tensor Cores $\mathbb{P}_{TC}$ moves the ridge point significantly to the right, making it harder to saturate compute resource. Under the condition ${I}^*_{CU} < I_{CU}^{(t)} < I_{TC}^{(t)} < {I}^*_{TC}$, the workload becomes memory-bound, and the performance ratio is formulated as:

\begin{equation} \label{eq.CB_to_MB_compare}
\begin{aligned}
    \frac{P_{TC,actual}^{(t)}}{P_{CU,actual}^{(t)}} 
    &= \frac{\frac{\mathbb{S}}{\alpha} \cdot \mathbb{B}\cdot t \cdot \frac{\alpha}{\mathbb{S}} \cdot \frac{{K}}{{D}}}{\mathbb{P}_{CU}} \\
    &> \frac{\frac{\mathbb{S}}{\alpha} \cdot \mathbb{B}\cdot t \cdot \frac{\alpha}{\mathbb{S}} \cdot \frac{{K}}{{D}}}{\mathbb{B}\cdot t \cdot \frac{{K}}{{D}}} \\
    &= 1
\end{aligned}
\end{equation}

This scenario highlights where Tensor Cores consistently \textbf{outperform} the baseline.
Figure~\ref{fig.case3} illustrates that the effective stencil performance on Tensor Cores reaches the hypothetical throughput of CUDA Cores with unlimited compute resource ${P'}_{CU}^{(t)}$. In effect, introducing Tensor Cores in this scenario \textit{breaks the performance ceiling} of CUDA Cores.

\textbf{(4) Compute-bound on CUDA Cores $\rightarrow$ Compute-bound on Tensor Cores.}

When implementations on both units are compute-bound, performance no longer scales with arithmetic intensity. Instead, the attainable throughput is limited by the peak compute throughput of the respective execution units. The ratio is given by:

\begin{equation}\label{eq.performance_under_sce4}
\frac{P_{TC,actual}^{(t)}}{P_{CU,actual}^{(t)}} = \frac{\frac{\mathbb{S}}{\alpha} \cdot \mathbb{P}_{TC}}{\mathbb{P}_{CU}}
\end{equation}

In this scenario, acceleration is \textbf{conditional rather than guaranteed}. Achieving performance gains requires a speedup ratio greater than 1. Since the hardware peak performances $\mathbb{P}_{CU}$, $\mathbb{P}_{TC}$ and sparse factor $\mathbb{S}$ are fixed for a given transformation on a specific hardware, the redundancy factor $\alpha$ becomes the decisive determinant. Consequently, effective acceleration necessitates:

\begin{equation} \label{eq.alpha_conditon}
    \alpha < \mathbb{S} \cdot \frac{\mathbb{P}_{TC}}{\mathbb{P}_{CU}}
\end{equation}

As formalized in Equation~\ref{eq.alpha}, $\alpha$ increases with the growth of fusion depth $t$. Taking the box-shaped stencil defined in Equation~\ref{eq.alpha_box} as an example, the redundancy factor is $\alpha_{box} = \frac{(2r \cdot t+1)^d}{t \cdot (2r+1)^d}$, exhibiting a polynomial growth rate of $O(t^{(d-1)})$. This scaling behavior implies that an aggressive fusion strategy rapidly inflates the redundancy factor $\alpha$, thereby diminishing the effective acceleration provided by Tensor Cores. Consequently, a careful selection of the fusion step $t$ is critical.

Moreover, leveraging the arithmetic intensity definitions in Equation~\ref{eq.cuda_t_ai} and Equation~\ref{eq.tc_t_ai}, the scaling factor $\frac{\mathbb{S}}{\alpha}$ can be reformulated as $\frac{{I}_{CU}^{(t)}}{{I}_{TC}^{(t)}}$. This leads to the relationship 
${P}_{TC,actual}^{(t)} = \frac{{I}_{CU}^{(t)}}{{I}_{TC}^{(t)}} \cdot P_{TC}^{(t)}$. Graphically, this condition delineates the profitable region, illustrated in dark green in Figure~\ref{fig.case4}. The union of this area and the light green region from Scenario 3 collectively defines the \textit{sweet spot} for Tensor Core acceleration.

\begin{figure}
    \centering
    \includegraphics[width=0.9\linewidth]{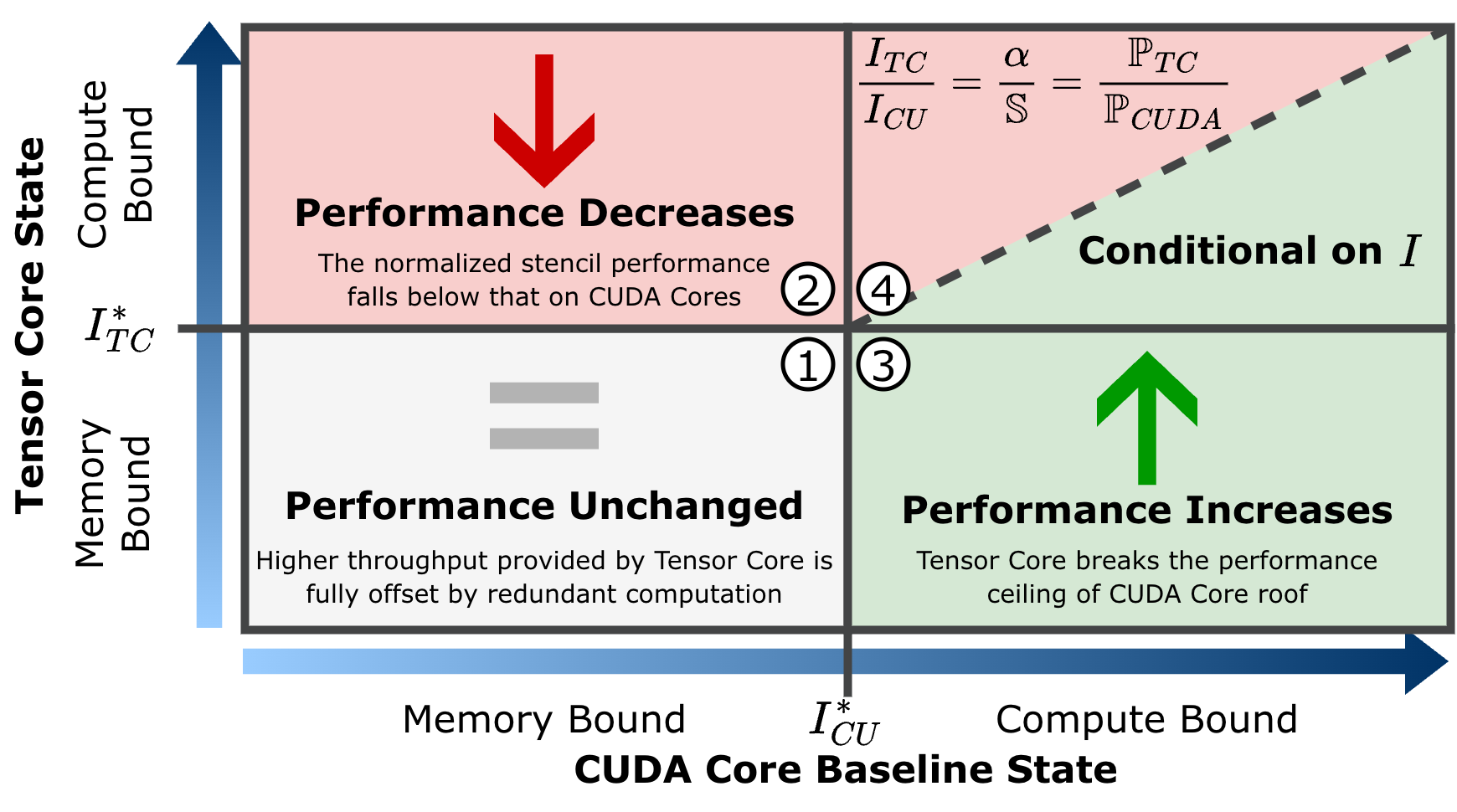}
    \caption{Illustration of Performance Criteria Utilizing Tensor Cores for Stencil Computations.}
    \label{fig.perf_class}
\end{figure}

\textbf{Summary.} 
Tensor Cores can indeed accelerate stencil computations, but their effectiveness is conditional. Figure~\ref{fig.perf_class} synthesizes the performance implications across four scenarios, offering a unified framework that resolves the apparent contradiction highlighted in \S~\ref{sec.intro}. Specifically, the figure (Scenarios 1 and 2) illustrates that when the stencil computation is memory-bound on CUDA Cores, utilizing Tensor Cores yields no benefit or even degrades performance. This observation provides a theoretical basis for the conventional view that Tensor Cores are not suited for stencils. Conversely, Scenarios 3 and 4 demonstrate that for compute-bound stencils, Tensor Cores unlock substantial performance gains, particularly within the \textit{sweet spot} (green region). This finding validates the empirical success reported in prior works such as TCStencil and ConvStencil. Ultimately, our model bridges the gap between conventional wisdom and recent empirical breakthroughs.

\subsection{Stencil is NOT Always Memory-Bound} \label{sec.not_memory_bound}

We have demonstrated that Tensor Cores provide performance boosts primarily when the workload saturates the compute resource of CUDA Cores (i.e., \textit{sweet spot}). While naive stencil computations typically exhibit low arithmetic intensity, techniques such as temporal fusion are widely adopted to alleviate memory pressure. Consequently, optimized stencil implementations frequently transition into the compute-bound region, depending on factors such as stencil shape, data type, fusion depth, and hardware specifications.

Motivated by this insight, we examine whether temporally fused stencil kernels have sufficient arithmetic intensity to benefit from Tensor Core acceleration. Our investigation proceeds in two phases: first, we employ theoretical analysis to identify the bottleneck migration; second, we verify these predictions with empirical results.

\begin{figure}[t]
    \centering
    \begin{subfigure}[b]{0.49\linewidth}
        \includegraphics[width=\linewidth]{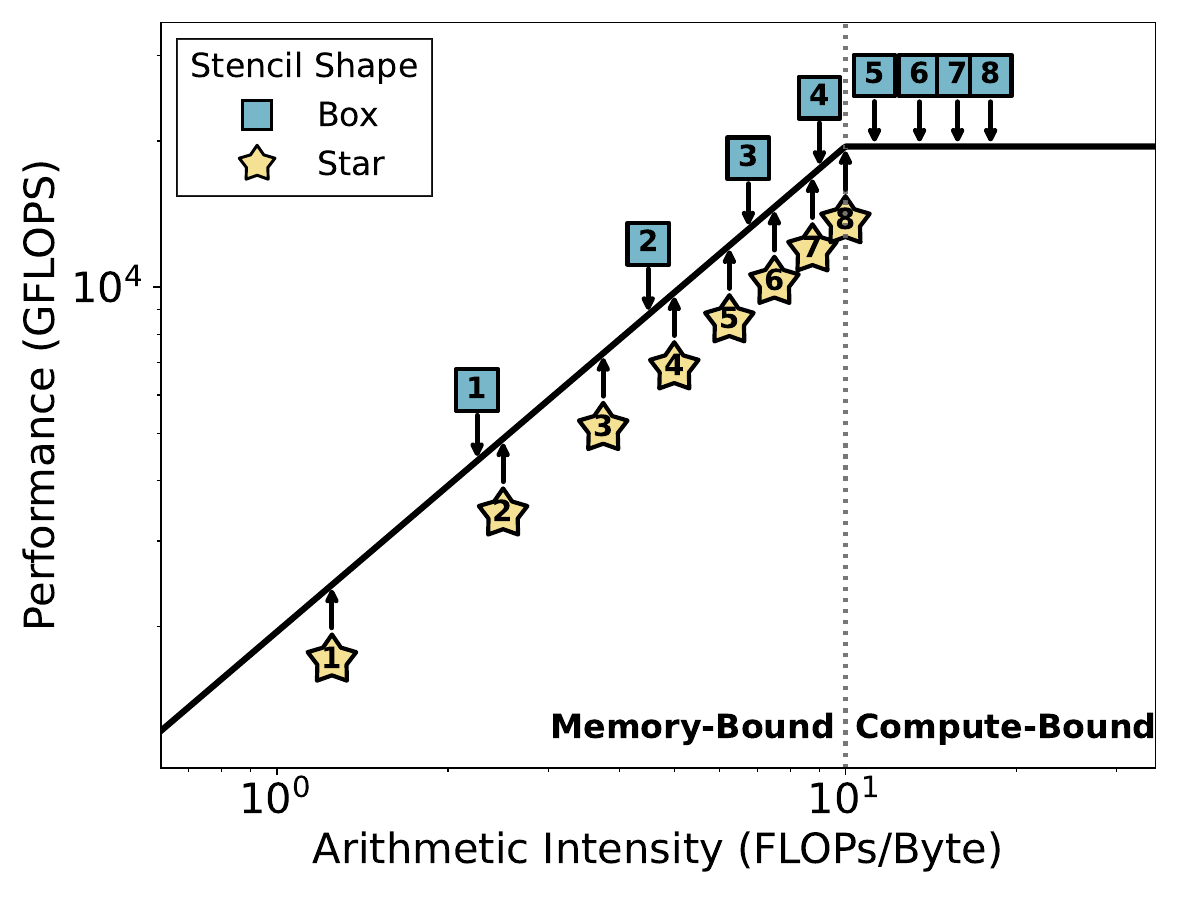}
        \caption{2D1R}
        \label{fig.roofline_2d1r}
    \end{subfigure}
    \begin{subfigure}[b]{0.49\linewidth}
        \includegraphics[width=\linewidth]{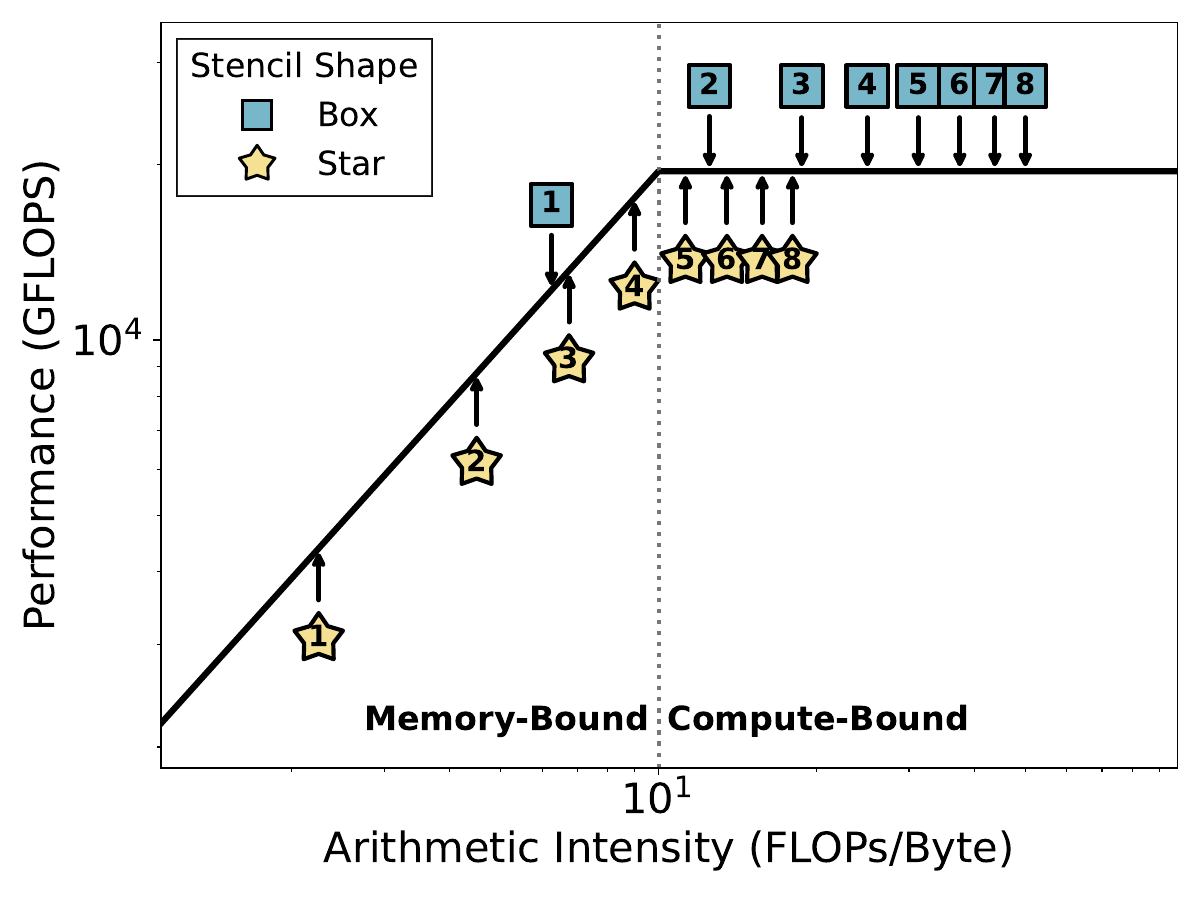}
        \caption{2D2R}
        \label{fig.roofline_2d2r}
    \end{subfigure}
    \begin{subfigure}[b]{0.49\linewidth}
        \includegraphics[width=\linewidth]{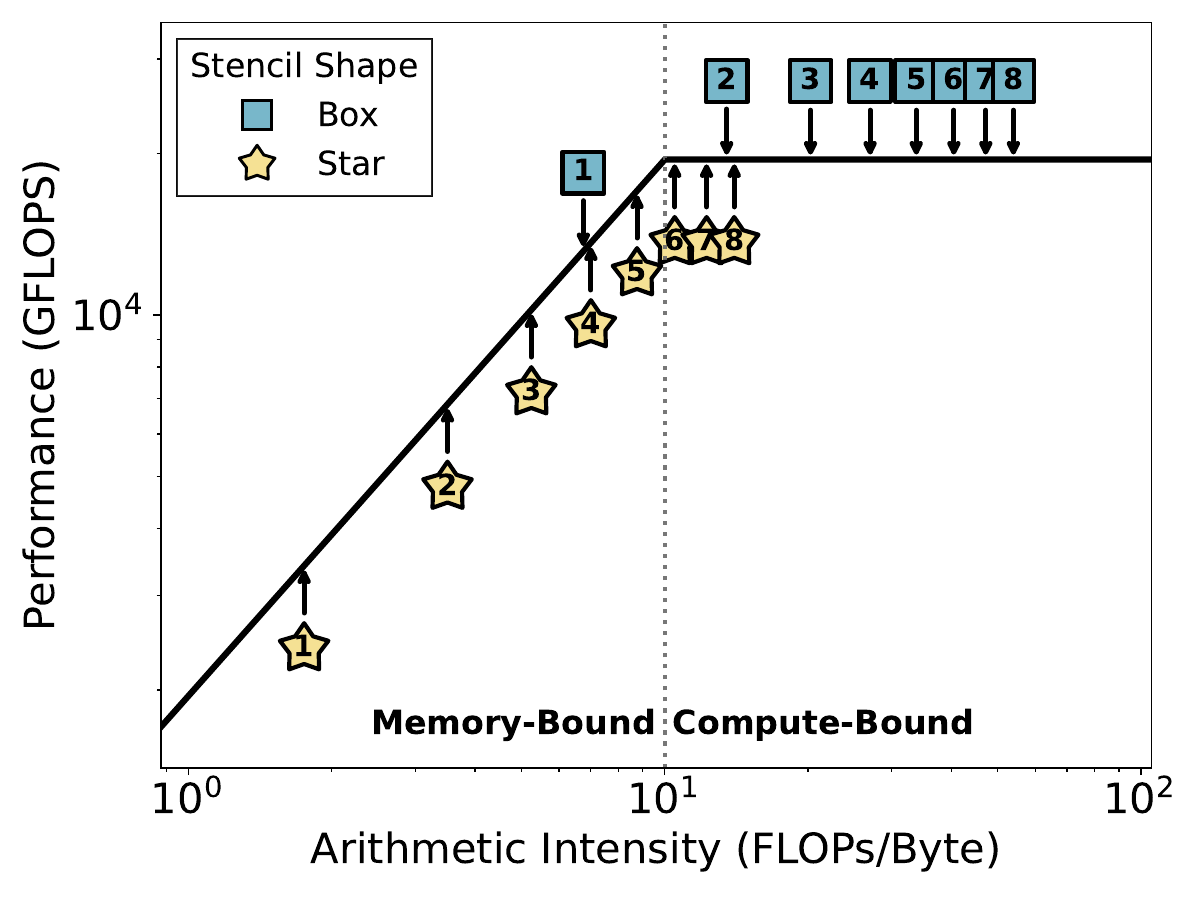}
        \caption{3D1R}
        \label{fig.roofline_3d1r}
    \end{subfigure}
    \begin{subfigure}[b]{0.49\linewidth}
        \includegraphics[width=\linewidth]{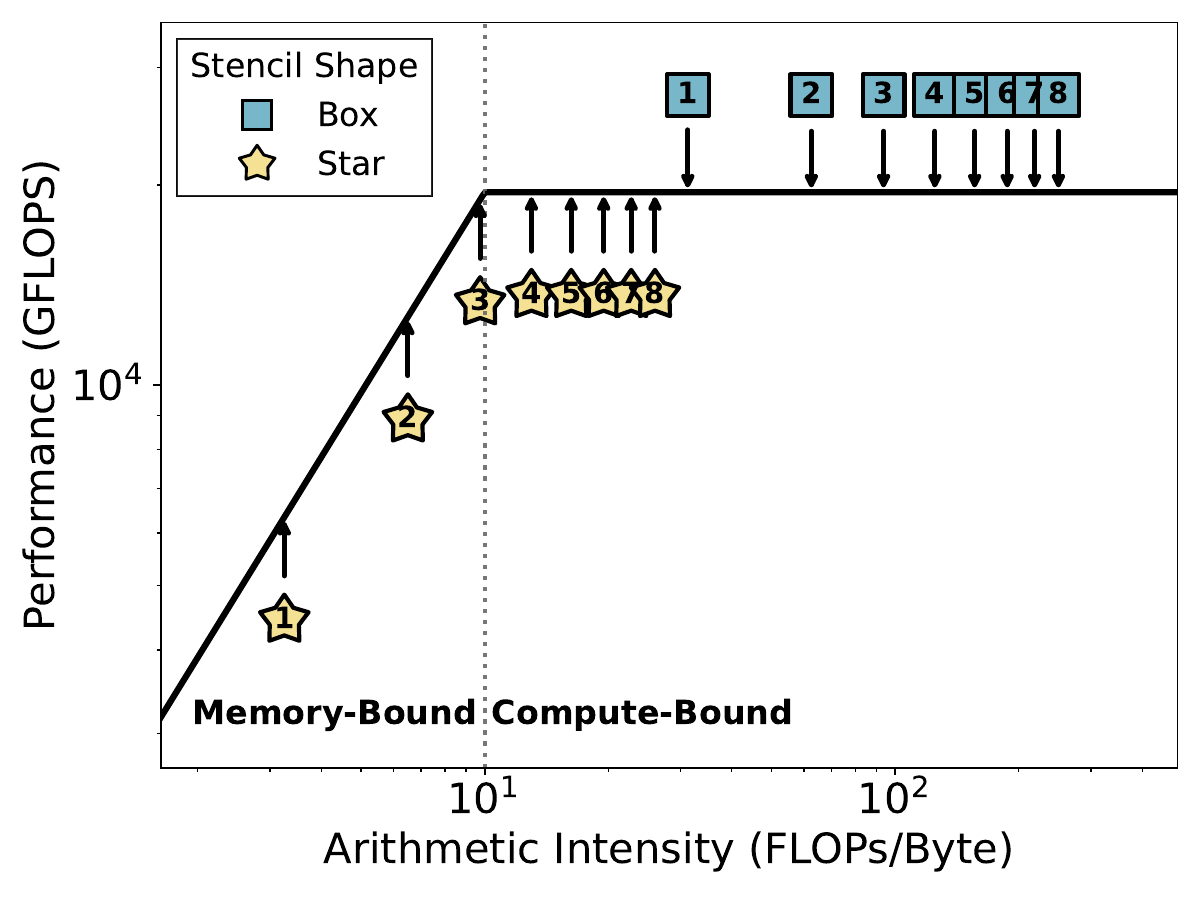}
        \caption{3D2R}
        \label{fig.roofline_3d2r}
    \end{subfigure}
    \caption{Problem Classification for Various Stencil Configurations on NVIDIA A100 GPU with Float Precision. The number in the marker denotes the temporal fuse steps.}
    \label{fig.theoretical_intensity}
\end{figure}

By applying Equation \ref{eq.cuda_t_ai}, we compare the theoretical arithmetic intensity of various representative stencil configurations against hardware limits, as shown in Figure \ref{fig.theoretical_intensity}.
Our analysis reveals a clear trend: most stencil problems transition into a compute-bound region after applying sufficient temporal fusion. Generally, kernels with higher dimensionality or larger stencil radius require fewer fusion steps to reach this transition. Notably, inherently intensive kernels, such as the Box-3D2R, operate in the compute-bound region even without fusion.

\begin{figure}[t]
\centering
\begin{subfigure}[b]{0.49\linewidth}
    \centering
    \includegraphics[width=\linewidth]{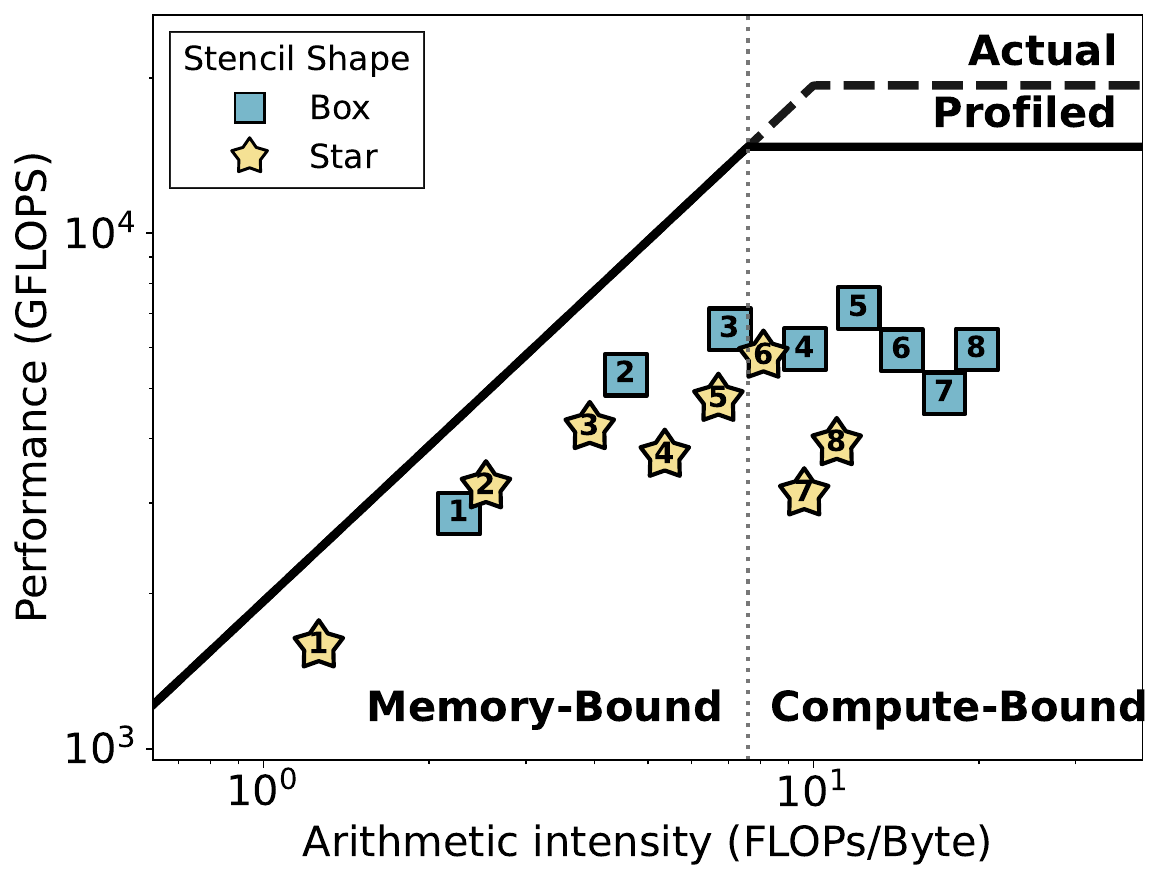}
    \caption{\texttt{float}}
    \label{fig:ebisu_fp32}
\end{subfigure}
\hfill
\begin{subfigure}[b]{0.49\linewidth}
    \centering
    \includegraphics[width=\linewidth]{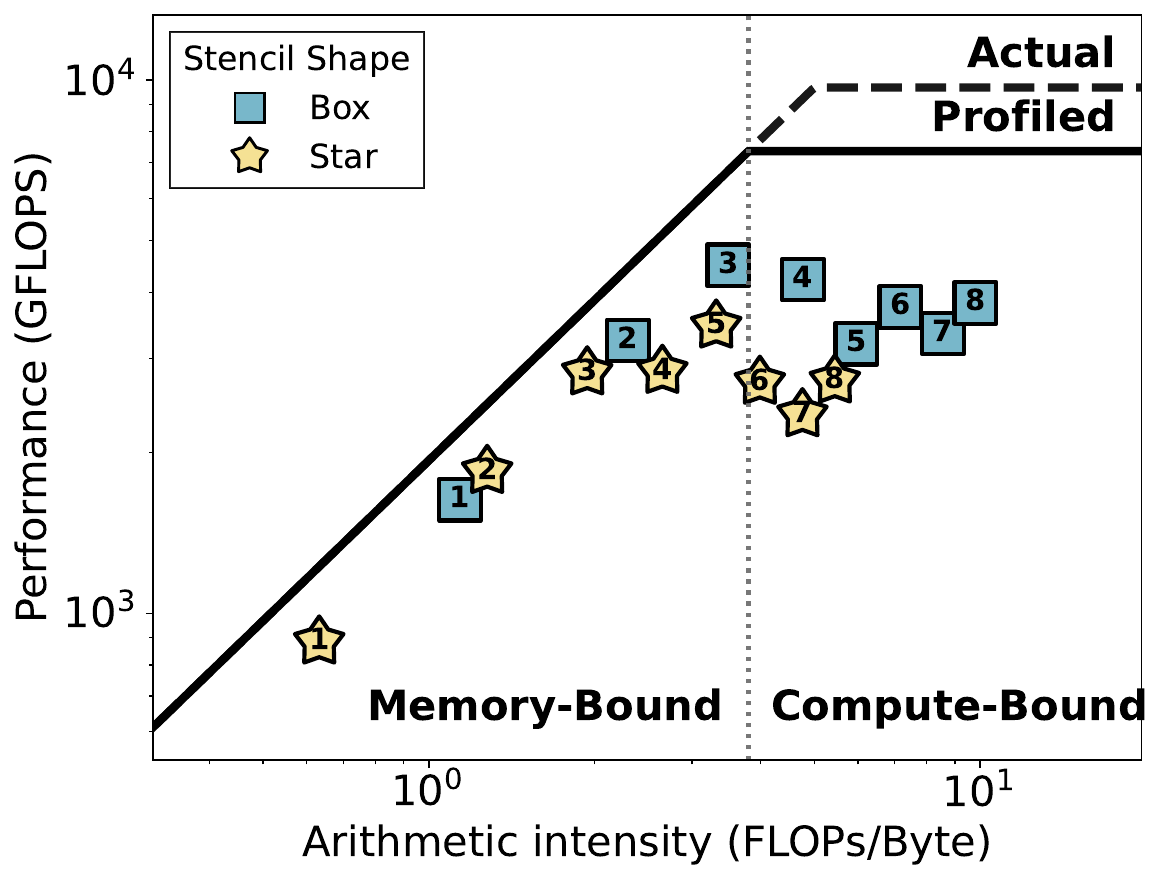}
    \caption{\texttt{double}}
    \label{fig:ebisu_fp64}
\end{subfigure}
\caption{The Roofline Chart Profiled from EBISU Implementation for 2D Stencil with Radius $r=1$ on A100 GPU. The number in the marker denotes the temporal fuse steps.}
\label{fig.ebisu_perf}
\end{figure}

Our empirical validation leverages EBISU~\cite{zhang2023revisiting}, a representative SOTA stencil framework. We conduct profiling on an NVIDIA A100 GPU, covering star- and box-shaped stencils (\texttt{float} and \texttt{double}) across fusion depths from 1 to 8. Consistent with our theoretical analysis, the results demonstrate that sufficient fusion depth shifts stencils into the compute-bound region. For instance, box stencils transition at $t=3$, while star stencils, with lower inherent intensity, transition at $t=5$.

It is worth noting that the empirically observed transition points appear slightly earlier than theoretical predictions. This discrepancy arises because the GPU clock is locked at a lower frequency during profiling to ensure stability and reproducibility of the results. This frequency locking lowers the effective compute ceiling, as reflected in Figure \ref{fig.theoretical_intensity}, thereby allowing the kernel to saturate compute resource at a shallower fusion depth. When using the actual peak performance, the theoretical and empirical transition points align closely.

These results effectively refute the prevailing belief that stencil computations are universally memory-bound. On the contrary, we demonstrate that SOTA stencil implementations, especially those leveraging temporal fusion, routinely saturate the compute resource of modern GPUs. This paradigm shift presents a compelling opportunity to harness Tensor Cores, utilizing their superior throughput to push performance further. 
This insight aligns with both the performance trends observed in recent studies and our analytical derivations in \S\ref{sec.analysis}.

\subsection{Further Acceleration with Sparse Hardware} \label{sec:sparse_further_accelerate}

\begin{figure}[t]
  \includegraphics[width=.95\linewidth]{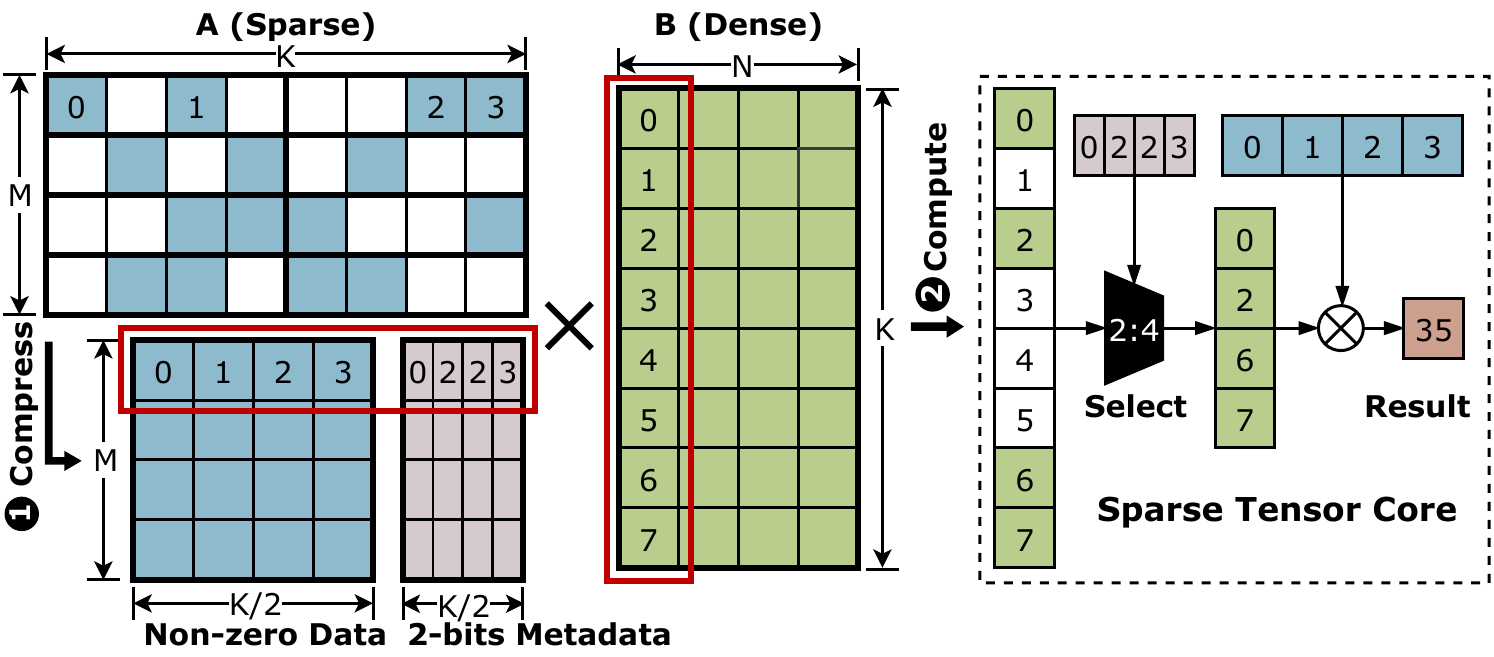} 
  \caption{2:4 Structured Sparse Format and its Compressed Representation for Sparse Tensor Cores.}
  \label{fig.sptc}
\end{figure}

As discussed in \S \ref{sec:sparse_in_transformation}, adapting stencil computations to Tensor Cores necessitates zero-padding to satisfy fixed operand size constraints. This introduces computational redundancy, as operations on padded zeros do not contribute to the final result. Fortunately, modern GPU architectures have augmented their matrix ALUs with native hardware support for sparsity. By skipping operations on zero-valued elements, the effective compute throughput can be significantly improved.

Specifically, NVIDIA integrates Sparse Tensor Cores (SpTCs)~\cite{amperewhitepaper} to accelerate sparse matrix operations. SpTCs rely on a 2:4 structured sparsity constraint (Figure~\ref{fig.sptc}), where each block of four elements contains a maximum of two non-zeros. This structure permits a compressed representation comprising packed values and 2-bit positional metadata. By skipping invalid elements based on this metadata, SpTCs deliver a 2$\times$ speedup compared to their dense counterparts. To leverage this throughput for stencil computations, recent studies like SparStencil~\cite{li2025sparstencil} and SPIDER~\cite{SPIDER} employ methods such as \textit{Strided Swapping} to adapt kernels into SpTC-compatible formats.

Adapting onto SpTC leaves the theoretical arithmetic intensity (Equation~\ref{eq.tc_t_ai}) unchanged (${I}^{(t)}_{\text{SpTC}} = {I}^{(t)}_{\text{TC}}$). Letting $\mathbb{P}_{SpTC}$ denote the peak throughput of SpTC, the resulting performance model is expressed as:

\begin{equation}
\begin{aligned}
    {I}^{(t)}_{\text{SpTC}} &= t \cdot \frac{\alpha}{\mathbb{S}} \cdot \frac{{K}}{{D}} \\
    {P}_{SpTC}^{(t)} &= \min(\mathbb{P}_{SpTC}, \mathbb{B}\cdot {I}^{(t)}_{SpTC}) \\
    {P}_{SpTC,actual}^{(t)} &= \frac{\mathbb{S}}{\alpha} \min(\mathbb{P}_{SpTC}, \mathbb{B}\cdot {I}^{(t)}_{SpTC})
\end{aligned}
\end{equation}

\begin{figure}
    \centering
    \includegraphics[width=0.93\linewidth]{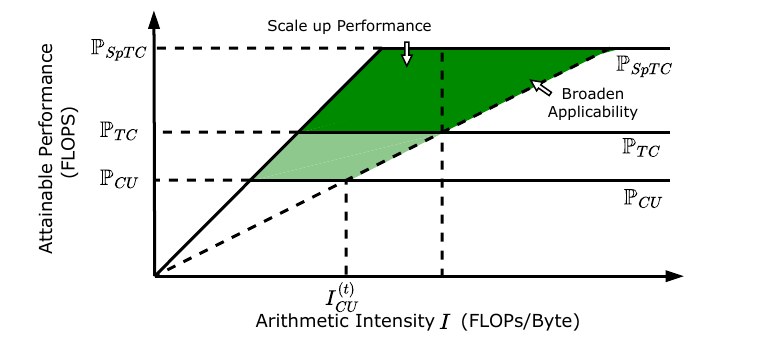}
    \caption{Sweet Spot for SpTC Acceleration.}
    \label{fig.sptc_sweet_zone}
\end{figure}

\begin{figure}
    \centering
    \includegraphics[width=0.9\linewidth]{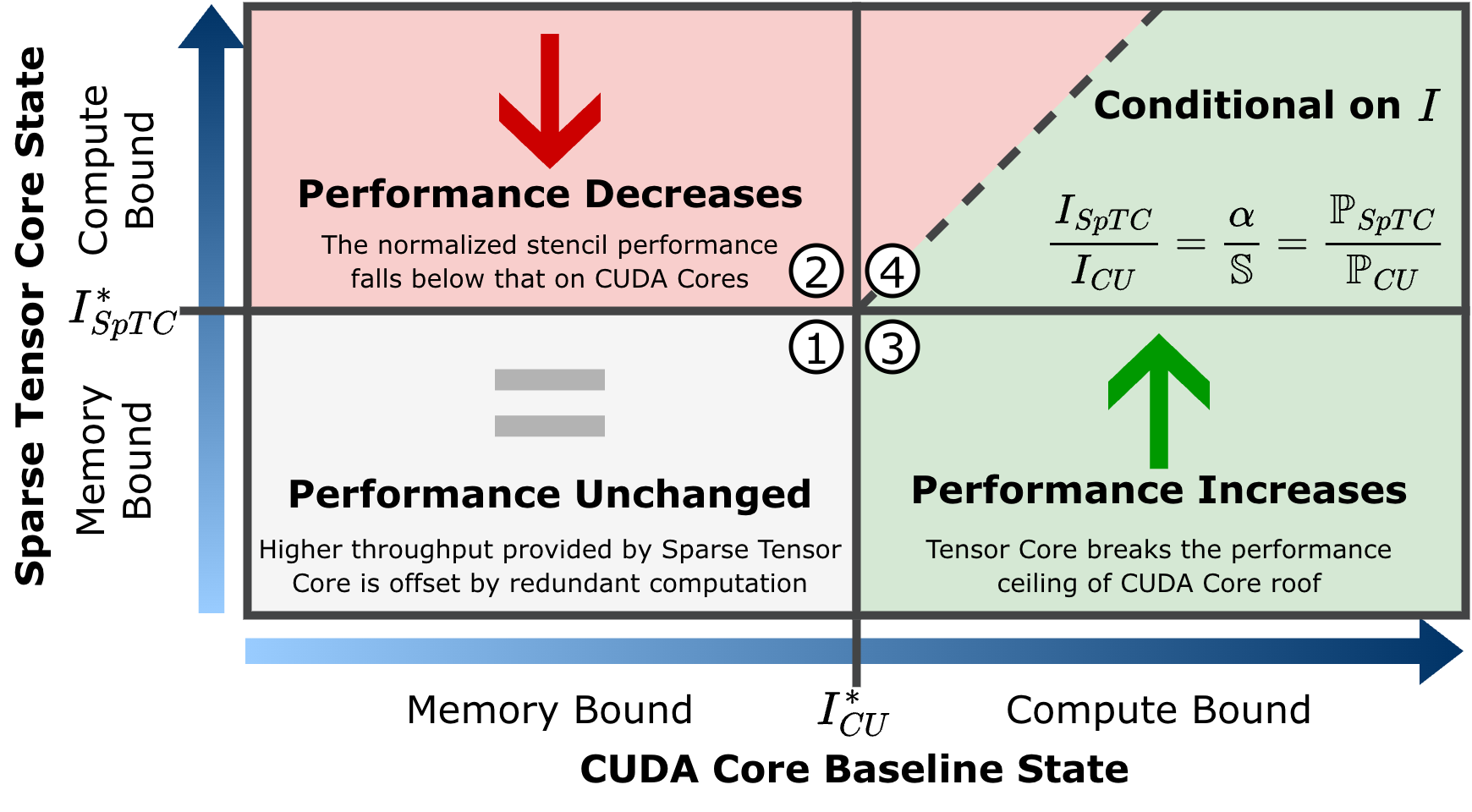}
    \caption{Illustration of Performance Criteria Utilizing Sparse Tensor Cores for Stencil Computations.}
    \label{fig.perf_class_sptc}
\end{figure}

Leveraging SpTCs yields a twofold benefit. First, it boosts the peak throughput for existing compute-bound workloads. By raising the hardware ceiling from $\mathbb{P}_{TC}$ to $\mathbb{P}_{SpTC}$, SpTCs directly scale up the performance of stencils that are already optimized on Tensor Cores, as shown in Figure \ref{fig.sptc_sweet_zone}. Second, it broadens the applicability of Tensor Core acceleration. On dense Tensor Cores, the throughput gain for stencils with aggressive temporal fusion is often offset by the overhead of redundant computations ($\frac{\alpha}{\mathbb{S}}$), rendering them slower than CUDA Core baselines. SpTCs mitigate this issue: their 2$\times$ throughput is sufficient to amortize the redundancy, converting previously inefficient configurations into viable candidates for acceleration. This effectively expands the \textit{sweet spot} for specialized hardware, as demonstrated in Figure~\ref{fig.perf_class_sptc}.

\section{Evaluation}

\begin{table*}[t]
\centering
\begin{tabular}{c cc c cc c ccc ccc}
\toprule
\multirow{2.5}{*}{} & \multirow{2.5}{*}{Baseline} & \multirow{2.5}{*}{\makecell{Stencil\\Pattern}} & \multirow{2.5}{*}{\makecell{Fusion\\Depth}} & \multirow{2.5}{*}{$\alpha$} & \multirow{2.5}{*}{$\mathbb{S}$} & \multirow{2.5}{*}{\makecell{Data\\ Type}} & \multicolumn{3}{c}{\textbf{Analytical}} & \multicolumn{3}{c}{\textbf{Experimental}} \\
\cmidrule(lr){8-10} \cmidrule(lr){11-13}
& & & & & & & $C$ & $M$ & $I$ & $C$ ($\Delta$) & $M $ ($\Delta$) & $I$ ($\Delta$) \\
\midrule
1 & \multirow{4}{*}{EBISU}  & Box-2D1R & 3 & / & / & double & 54  & 16 & 3.38  & 55.78 (3.30\%)  & 15.95 (-0.30\%) & 3.50 (3.61\%) \\
                        2 & & Box-2D3R & 1 & / & / & double & 98  & 16 & 6.12  & 100.65 (2.70\%) & 15.92 (-0.50\%) & 6.32 (3.21\%) \\
                        3 & & Box-2D1R & 7 & / & / & float  & 126 & 8  & 15.75 & 137.35 (9.01\%) & 7.93 (-0.90\%)  & 17.32 (10.00\%) \\
                        4 & & Box-2D7R & 1 & / & / & float  & 450 & 8  & 56.25 & 484.51 (7.61\%) & 7.91 (-1.09\%)  & 61.23 (8.85\%) \\
\midrule
5 & \multirow{4}{*}{ConvStencil} & Box-2D1R & 3 & 1.81 & 0.5 & \multirow{2}{*}{double} & \multirow{2}{*}{196} & \multirow{2}{*}{16} & \multirow{2}{*}{12.25} & \multirow{2}{*}{208.00 (6.12\%)} & \multirow{2}{*}{15.91 (-0.56\%)} & \multirow{2}{*}{13.07 (6.73\%)} \\
                             6 & & Box-2D3R & 1 & 1 & 0.5 &&&&&&& \\
                             \cmidrule(lr){7-13}
                             7 & & Box-2D1R & 7 & 3.57 & 0.5 & \multirow{2}{*}{float} & \multirow{2}{*}{900} & \multirow{2}{*}{8} & \multirow{2}{*}{112.50} & \multirow{2}{*}{928.00 (3.11\%)} & \multirow{2}{*}{8.27 (3.36\%)} & \multirow{2}{*}{112.23 (-0.24\%)} \\
                             8 & & Box-2D7R & 1 & 1 & 0.5 &&&&&&& \\
\midrule
9 & \multirow{2}{*}{SPIDER}      & Box-2D1R & 7 & 3.57 & 0.47 & \multirow{2}{*}{float}  & \multirow{2}{*}{960} & \multirow{2}{*}{8} & \multirow{2}{*}{120.00} & \multirow{2}{*}{960.00 (0.00\%)} & \multirow{2}{*}{7.89 (-1.35\%)}  & \multirow{2}{*}{121.64 (1.37\%)} \\
                             10 & & Box-2D7R & 1 & 1 & 0.47 &&&&&&& \\
\bottomrule
\end{tabular}
\caption{Comparison of Analytical and Experimental Metrics across Different Baselines. For CUDA Core implementation, factor $\alpha$ and $\mathbb{S}$ are not involved in its formulation.}
\label{tab.formulation_compare}
\end{table*}

\subsection{Evaluation Setup}

\textbf{Platforms.} Our evaluation platform utilizes two Intel Xeon Platinum 8558P CPUs, 512GB DDR5 memory, and NVIDIA A100-80GB PCIe GPU. The system operates on Ubuntu 22.04 LTS with CUDA 12.8 and cuDNN 9.8.0.

\textbf{Baselines.}
In this section, we consider a comprehensive set of baselines, including cuDNN~\cite{chetlur2014cudnn}, DRStencil~\cite{you2021drstencil} and EBISU~\cite{zhang2023revisiting} for CUDA Core architectures, as well as TCStencil~\cite{TCStencil}, ConvStencil~\cite{chen2024convstencil}, LoRAStenicl~\cite{zhang2024lorastencil} and SPIDER~\cite{SPIDER} for Tensor Core architectures.

To verify the correctness of our performance formulation and analytical criteria, we specifically select SOTA implementations on CUDA Cores (EBISU), Tensor Cores (ConvStencil and SPIDER) as representative methods. This selection is grounded in the performance analysis presented in \S \ref{sec:exp.overall}, where these methods are shown to deliver SOTA performance in their respective categories. By targeting the most competitive implementations, we ensure the rigor and generality of our evaluation.

It should be noted that we adapt the baselines in two aspects. First, we extend the codebases to support unimplemented stencil shapes to enable case-by-case comparison. For instance, we implement Box-2D7R support for EBISU. Second, we unify the handling of stencil kernels to ensure a fair comparison. While some baselines originally compile stencil parameters as constant literals, we modify all implementations to support dynamic kernel values at runtime.

\textbf{Benchmarks.} We evaluate our approach using various stencil shapes ranging from 2D to 3D. For 2D stencils, we employ a domain size of $10240\times10240$ and evaluate both Box and Star shapes (radius $r\in\{1,3,7\}$). For 3D stencils, we utilize a domain size of $1024\times1024\times1024$, covering both Box and Star shapes with $r=1$. The radius is selected to align with the dimension requirement of Tensor Cores.

\subsection{Verification of Proposed Formulations} \label{sec.eval_verify}

We assess the accuracy of our model by comparing its analytical predictions with empirical data extracted from NVIDIA Nsight Compute (\texttt{ncu}). Specifically, we estimate the analytical metrics for computation $C$, memory traffic $M$, and arithmetic intensity $I$ using Equation~\ref{eq.cuda_t_ai} and ~\ref{eq.tc_t_ai}. The redundancy factor $\alpha$ is derived from Equation~\ref{eq.alpha_box}, while $\mathbb{S}$ remains constant for each implementation. For validation, we profile the computing kernels using the \texttt{ncu} profiler, correlating \textit{achieved work} and \textit{achieved traffic} with the modeled $C$ and $M$. A detailed comparison across different baselines and patterns is shown in Table~\ref{tab.formulation_compare}.

\subsubsection{Memory Traffic Analysis ($M$)}
First, we examine the behavior of memory traffic. As detailed in Table~\ref{tab.formulation_compare}, the experimental values for $M$ remain nearly constant across varying temporal fusion depth~$t$. For instance, in the EBISU baseline, comparing the data in Rows 1 and 2 (as well as Rows 3 and 4) reveals that increasing $t$ induces negligible fluctuations in memory traffic. The experimental results confirm that global memory traffic is dominated by compulsory data movement, as intermediate updates are effectively handled via on-chip memory, thereby decoupling global memory traffic from the temporal depth and validating our formulation.

\subsubsection{Computational Operation Analysis ($C$)} 
In terms of computational operations, our model exhibits high predictive accuracy across different hardware backends. For the CUDA Core-based EBISU, the measured $C$ aligns closely with analytical predictions, confirming the precision of our fundamental operation counting logic. For Tensor Core-based methods (ConvStencil and SPIDER), we further analyze the computational redundancy introduced by hardware adaptation. Comparing Rows 7 and 9 (identical temporal depth $t$, different $\mathbb{S}$) confirms that the measured computation scales correctly with our estimated sparsity factor $\mathbb{S}$. Conversely, fixing $\mathbb{S}$ while varying $t$ (Rows 5 vs. 6, 7 vs. 8, and 9 vs. 10) reveals that the computational overhead follows the redundancy ratio $\alpha$ defined in Equation~\ref{eq.alpha_box}. Collectively, these results show that the total computation volume aligns with Equation~\ref{eq.compute_ops}, proving that our model accurately encapsulates both intrinsic arithmetic and the necessary padding overheads.

\begin{figure}[t]
\centering
\begin{subfigure}[b]{0.49\linewidth}
    \centering
    \includegraphics[width=\linewidth]{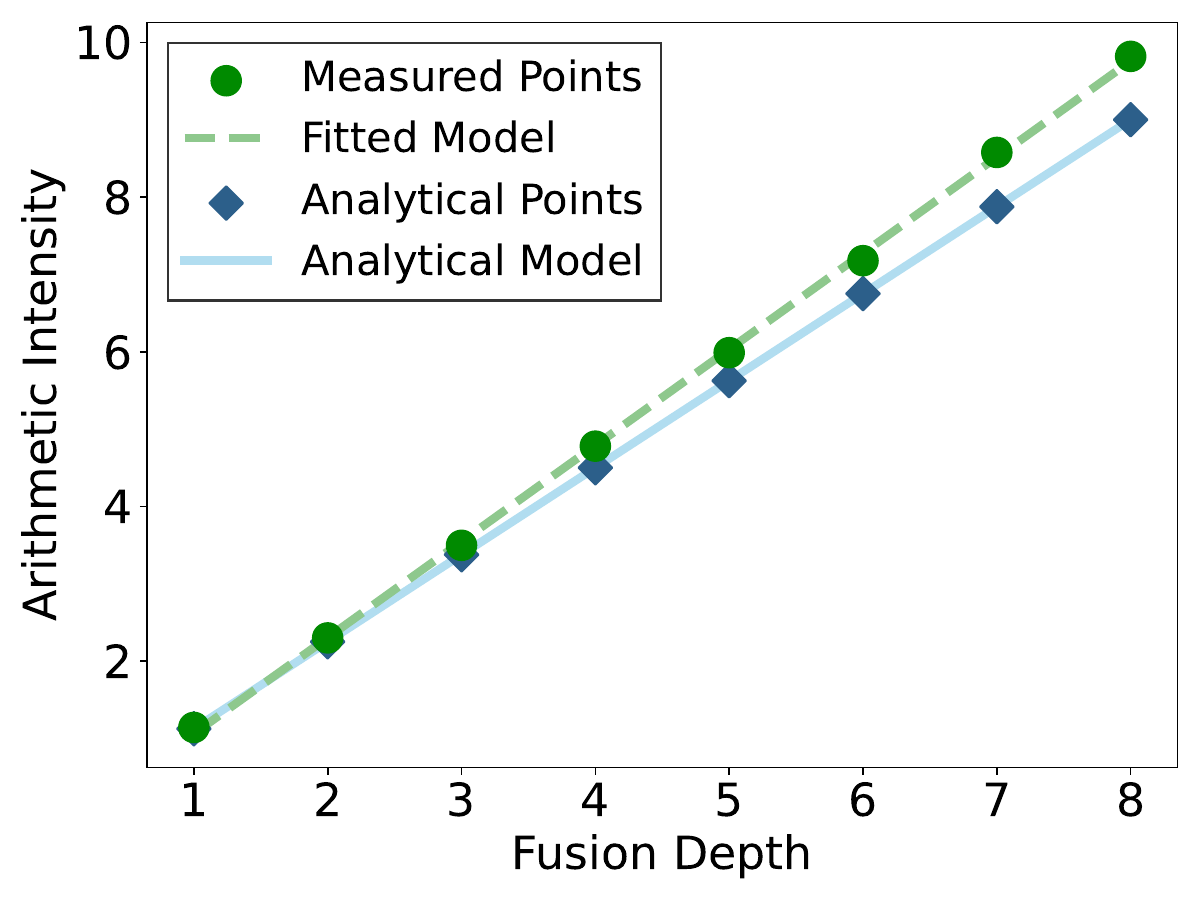}
    \caption{Box-2D1R}
    \label{fig:box_fp32}
\end{subfigure}
\hfill
\begin{subfigure}[b]{0.49\linewidth}
    \centering
    \includegraphics[width=\linewidth]{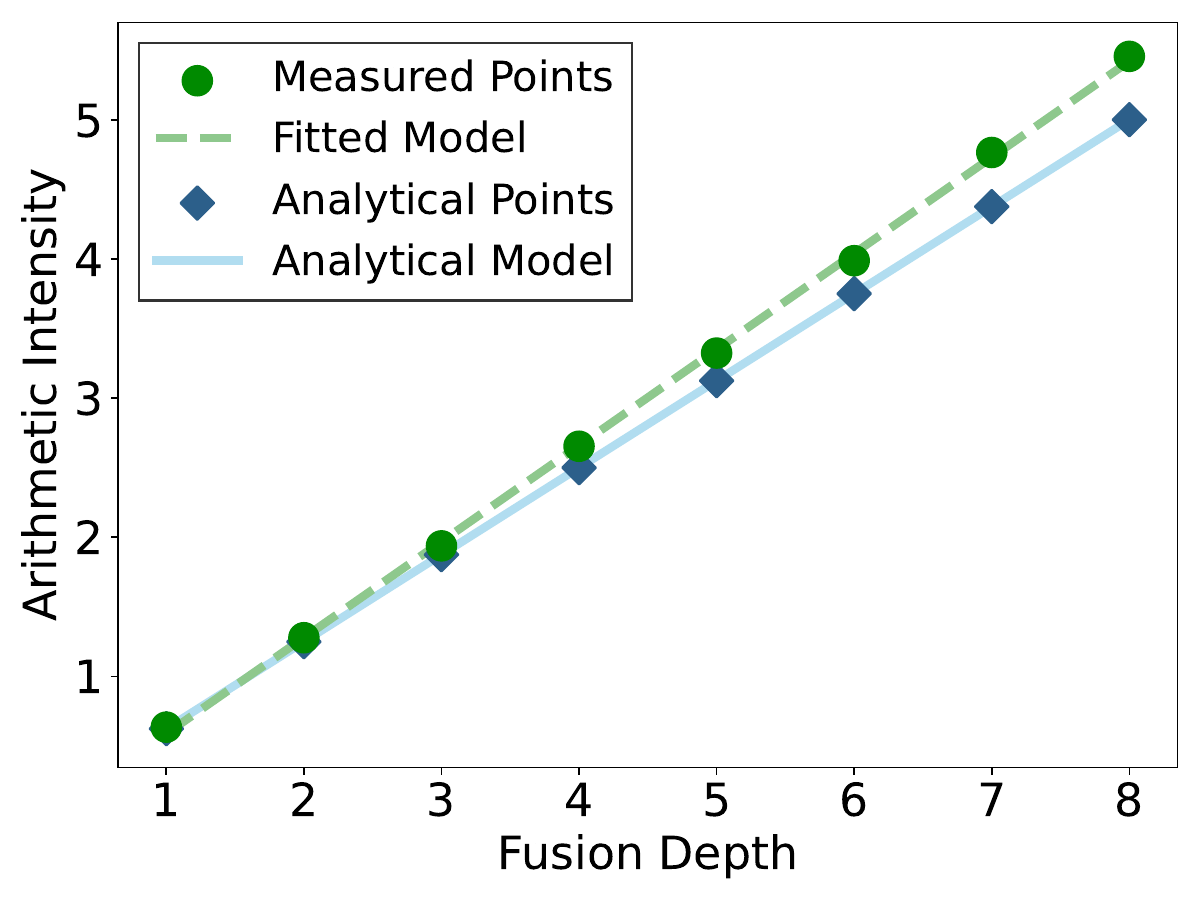}
    \caption{Star-2D1R}
    \label{fig:star_fp64}
\end{subfigure}
\caption{Arithmetic Intensity vs. Fusion Depth for CUDA Core Implementation with Double Precision.}
\label{fig.ai_vs_t}
\end{figure}

\begin{table*}[t]
    \centering
    \begin{tabular}{c ccc cccccc c c}
    \toprule
    Case &
    \makecell{Stencil\\Pattern} &
    \makecell{Fuse\\Depth} &
    \makecell{Data\\Type} & 
    \makecell{Baseline} & 
    \makecell{Arithmetic \\ Intensity} & 
    \makecell{Ridge \\ Point} & 
    \makecell{Bottleneck} & 
    \makecell{Performance\\(GStencils/sec)} &
    \makecell{Performance\\Change}&
    \makecell{Belonging\\Scenario}\\
    \midrule
    \multirow{2}{*}{\ding{182}} & \multirow{2}{*}{Box-2D1R} & \multirow{2}{*}{3} & \multirow{2}{*}{double} & EBISU      & 3.38   & 5  & Memory  & 260.90 & \multirow{2}{*}{$\downarrow$} &\multirow{2}{*}{Scenario 2}\\
    & &&& ConvStencil& 12.25   & 10 & Compute & 190.14 \\
    \midrule
    \multirow{2}{*}{\ding{183}} & \multirow{2}{*}{Box-2D3R} & \multirow{2}{*}{1} & \multirow{2}{*}{double} & EBISU      & 6.13   & 5  & Compute  & 64.05 &\multirow{2}{*}{$\approx$} & \multirow{2}{*}{Scenario 4}\\
    & &&& ConvStencil& 12.25   & 10 & Compute & 63.33\\
    \midrule
    \multirow{2}{*}{\ding{184}} & \multirow{2}{*}{Box-2D1R} & \multirow{2}{*}{7} & \multirow{2}{*}{float} & EBISU           & 15.75   & 10  & Compute   & 318.31 & \multirow{2}{*}{$\uparrow$} & \multirow{2}{*}{Scenario 3} \\
    & &&& SPIDER   & 120.00     & 161 & Memory    & 1002.94 \\
    \midrule
    \multirow{2}{*}{\ding{185}} & \multirow{2}{*}{Box-2D7R} & \multirow{2}{*}{1} & \multirow{2}{*}{float} & EBISU           & 21.58   & 10  & Compute & 50.35 &\multirow{2}{*}{$\uparrow$} & \multirow{2}{*}{Scenario 3} \\
    & &&& SPIDER   & 120.00     & 161 & Memory  & 143.28 \\
    \midrule
    \multirow{2}{*}{\ding{186}} & \multirow{2}{*}{Box-3D1R} & \multirow{2}{*}{3} & \multirow{2}{*}{double} & EBISU           & 10.13   & 5  & Compute & 37.74 &\multirow{2}{*}{$\downarrow$}& \multirow{2}{*}{Scenario 4} \\
    & &&& ConvStencil   & 85.75     & 10 & Compute  & 24.63 \\
    \midrule
    \multirow{2}{*}{\ding{187}} & \multirow{2}{*}{Box-3D1R} & \multirow{2}{*}{7} & \multirow{2}{*}{float} & EBISU           & 47.25   & 10  & Compute & 71.23 & \multirow{2}{*}{$\downarrow$} &\multirow{2}{*}{Scenario 4} \\
    & &&& SPIDER   & 1795.21     & 161 & Compute  & 51.13 \\
    \bottomrule
    \end{tabular}
    \caption{Comparison of Stencil Performance and Bottleneck Transitions across Representative Cases.}
    \label{tab.perf_compare}
\end{table*}

\subsubsection{Arithmetic Intensity Analysis ($I$)} 

Finally, we evaluate the arithmetic intensity $I = C/M$ to validate our analytical model. Overall, the experimental results closely match the analytical derivation across both CUDA Core and Tensor Core implementations.

To rigorously verify the correlation between arithmetic intensity and the temporal fusion depth $t$, we conducted an extensive experiment using the CUDA Core baseline, leveraging its flexibility in the configuration of fusion depth $t$. As illustrated in Figure~\ref{fig.ai_vs_t}, the results reveal a clear linear relationship between $I$ and $t$, providing strong empirical evidence for our performance model.

\subsubsection{Summary and Deviation Analysis} 

The evaluation results confirm that our analytical formulation captures the fundamental performance characteristics of the system, yielding accurate estimation for $C$, $M$, and $I$ across a diverse range of configurations. While the overall correlation is strong, we observe minor yet systematic deviations between the modeled and measured data, which reflect specific hardware behaviors excluded from our model.

Specifically, the measured values for $C$ consistently exceed analytical predictions. We attribute this slight inflation to overheads in specific implementation, primarily the redundant calculations in halo regions arising from thread block overlaps. These operations are captured by the hardware profiler but are excluded from our formulation. Conversely, the measured memory traffic $M$ is generally lower than predicted. This discrepancy arises because our analytical model assumes all data transfers originate from DRAM (compulsory misses). In practice, the L2 cache effectively filters memory requests, serving a portion of accesses and thereby reducing the actual off-chip traffic observed at the memory controller.

\subsection{Analytical Criteria Verification}

\label{sec.eval_criteria}

To assess the validity of our proposed criteria, we conducted an analysis of six representative cases (Cases \ding{182}--\ding{187}), by comparing the analytical conclusion and the empirical performance. These cases include all the cases currently available in SOTA stencil computations on Tensor Cores, encompassing the three distinct scenarios analyzed in \S\ref{sec.analysis}. We quantify the performance using GStencils/sec, the de facto standard metric for stencil throughput. The detailed configurations, bottleneck shifts, and performance results are summarized in Table~\ref{tab.perf_compare}.

\textbf{Memory-bound on CUDA Cores $\to$ Compute-bound on Tensor Cores.}
Case \ding{182} exemplifies this scenario. In the baseline implementation (EBISU), the performance is constrained by memory bandwidth (memory-bound). However, due to the redundancy factor $\alpha$ and the sparsity in padded matrices $\mathbb{S}$, the Tensor Core implementation (ConvStencil) is pushed into a compute-bound region.

Our analytical criteria, as derived in Equation~\ref{eq.MB_to_CB_compare}, predict that for this scenario, the actual performance is penalized by the computational ceiling of Tensor Cores, causing the effective compute throughput lower than the baseline. Experimental results corroborate this prediction: the introduction of Tensor Cores resulted in a 27.12\% performance degradation.

\textbf{Compute-bound on CUDA Cores $\to$ Memory-bound on Tensor Cores.}
Cases \ding{184} and \ding{185} fall into this category. In these cases, the baseline implementation is strictly constrained by the peak floating-point performance of CUDA Cores ($P_{CU}$), whereas the Tensor Core implementation (SPIDER) successfully alleviates this compute bottleneck.

Analytically, this scenario represents the ideal use case for acceleration, where the introduction of Tensor Cores breaks the computational ceiling of CUDA Cores, thus provides performance benefits, as derived in Equation~\ref{eq.CB_to_MB_compare}. Empirically, this is confirmed by substantial speedups of 7.73$\times$ and 6.64$\times$ for Case \ding{184} and \ding{185}, respectively.

\textbf{Compute-bound on CUDA Cores $\to$ Compute-bound on Tensor Cores.}
This scenario is the most intricate one, where performance gains are \textit{conditional rather than guaranteed}, as derived in Equation~\ref{eq.performance_under_sce4}. We use Cases \ding{183}, \ding{186}, and \ding{187} to verify if our \textit{sweet spot} formulation accurately delineates the boundary of performance benefits.

For Case \ding{186}, the redundancy factor is calculated as $\alpha \approx 1.81$. Given the hardware specifications of A100 GPU ($\mathbb{P}_{CU} = 9.7$ TFLOPS and $\mathbb{P}_{TC} = 19.5$ TFLOPS for double precision), this case fails to satisfy the condition defined in Equation~\ref{eq.alpha_conditon} ( $\alpha > \mathbb{S} \cdot \frac{\mathbb{P}_{TC}}{\mathbb{P}_{CU}}$). Consequently, it falls outside the \textit{sweet spot}, resulting in performance degradation. The experimental observation matches our prediction.

Similarly, for Case \ding{187}, the larger stencil dimension leads to a higher $\alpha$ value, again violating the \textit{sweet spot} condition. Both theoretical analysis and experimental data indicate a performance degradation.

In contrast, Case \ding{183} presents a boundary condition. According to Equation~\ref{eq.performance_under_sce4}, the ratio of the Tensor Core performance to the CUDA Core performance approaches 1. 
This suggests the case lies near the boundary of the \textit{sweet spot}, implying comparable performance between the two implementations. The experimental results confirm this, showing negligible performance difference.

\textbf{Summary.} Collectively, these experiments demonstrate the validity of our analytical model. By accurately predicting positive, negative, and neutral outcomes across diverse scenarios, our criteria prove effective in guiding the selection of the optimal execution unit for stencil computations.

\subsection{Benefits from Sparse Tensor Core}

\begin{table}[t]\small
    \centering
    \begin{tabular}{cccccc}
    \toprule
    \makecell{Baseline} & 
    \makecell{Arithmetic \\ Intensity} & 
    \makecell{Ridge \\ Point} & 
    \makecell{Bottleneck} & 
    \makecell{Performance\\(GStencils/sec)} \\
    \midrule
    SPIDER-Dense    & 120     & 81  & Compute   & 327.39  \\
    SPIDER-Sparse   & 120     & 161 & Memory    & 1002.94 \\
    \bottomrule
    \end{tabular}
    \caption{Comparison between Dense and Sparse Tensor Cores (Box-2D1R with $t$=7 for float precision).}
    \label{tab.2d7r_t1}
\end{table}

To validate the theoretical advantages of sparse hardware detailed in \S~\ref{sec:sparse_further_accelerate}, we compare the native SPIDER implementation utilizing Sparse Tensor Cores against its variant employing dense Tensor Cores. Table \ref{tab.2d7r_t1} presents the performance metrics for a Box-2D1R stencil with temporal fusion $t$=7. The results demonstrate that the introduction of SpTCs fundamentally alters the performance bottleneck, transitioning the workload from a compute-bound to a memory-bound region and delivering a remarkable 3.06$\times$ speedup over the dense baseline. This empirical evidence confirms that the enhanced compute throughput of SpTCs effectively shifts the hardware ridge point and broadens the acceleration scope for stencil computations, thereby elevating the attainable performance.

\subsection{Overall Performance Comparison} \label{sec:exp.overall}

\begin{figure}[t]
\begin{subfigure}[b]{0.95\linewidth}
    \centering
    \includegraphics[width=\linewidth]{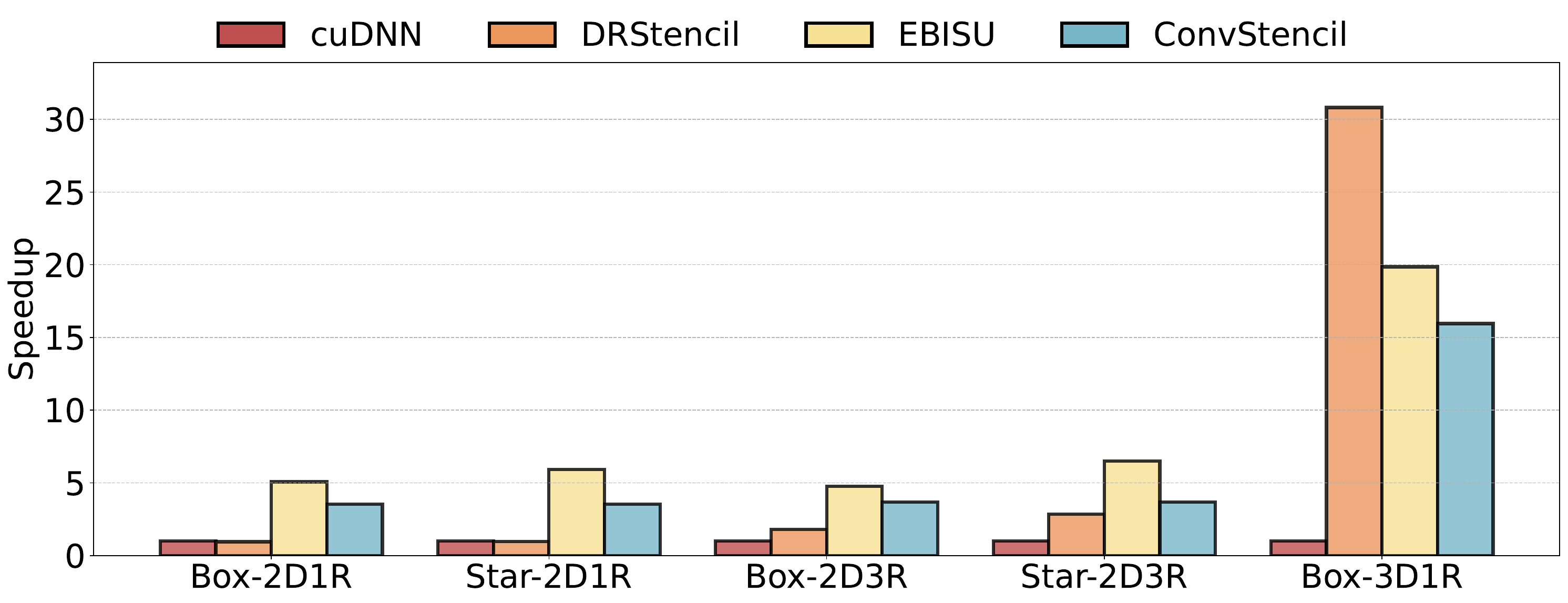}
    \caption{\texttt{double}}
    \label{fig:overall_fp64}
\end{subfigure}
\hfill
\begin{subfigure}[b]{0.95\linewidth}
    \centering
    \includegraphics[width=\linewidth]{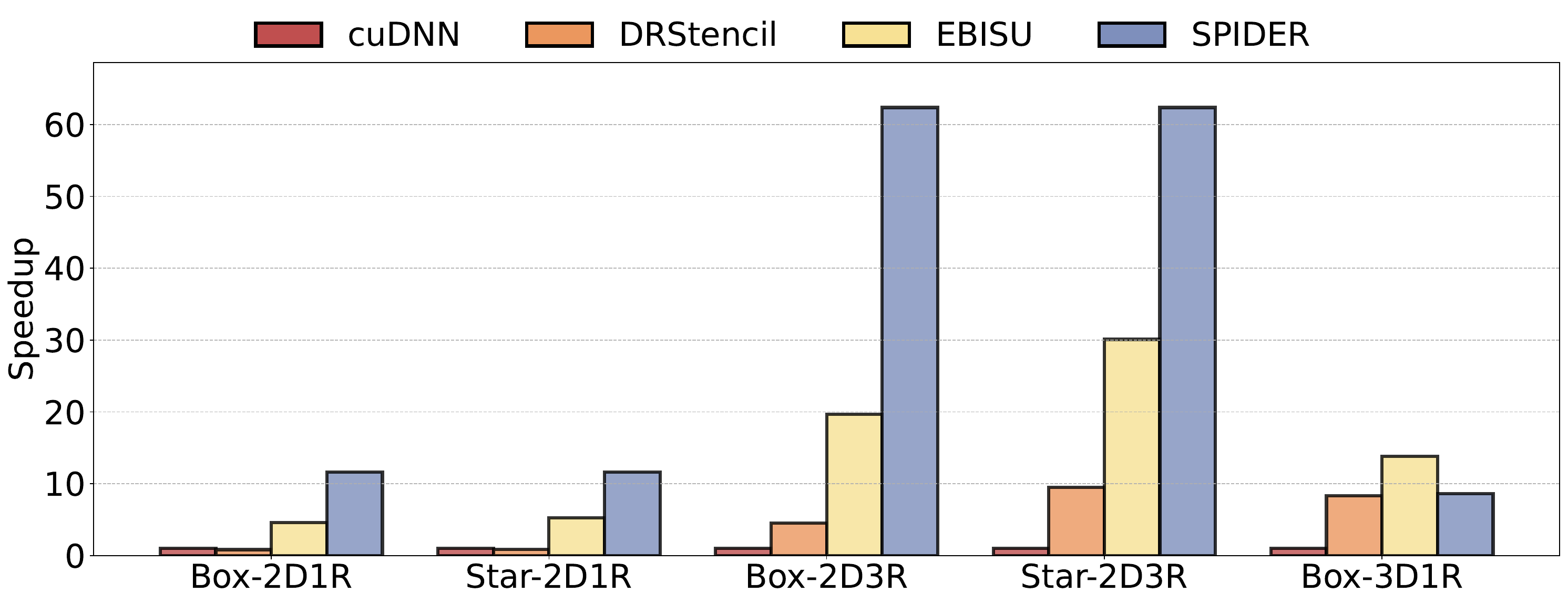}
    \caption{\texttt{float}}
    \label{fig:overall_fp32}
\end{subfigure}
\caption{Overall Performance Comparison.}
\label{fig.perf_compare_precision}
\end{figure}

Finally, we evaluate the overall stencil performance across the different data precisions supported by the baselines. For double precision (\texttt{double}), we benchmark cuDNN, DRStencil, EBISU, and ConvStencil. For single precision (\texttt{float}), we evaluate cuDNN, DRStencil, EBISU, and SPIDER. Two baselines are excluded from this analysis: TCStencil, which is limited to half-precision support, and LoRAStencil, which assumes symmetric stencil kernels and is therefore inapplicable to general-purpose scenarios. 
As shown in Figure \ref{fig.perf_compare_precision}, EBISU, ConvStencil, and SPIDER represent the SOTA implementations on CUDA Cores, Tensor Cores, and Sparse Tensor Cores, respectively. These results validate the selection of baselines used for the analysis in our evaluation.

\section{Related Work}

\textbf{Traditional Stencil Optimizations on General-Purpose Architectures.}
Traditional stencil optimizations on CPUs primarily focus on mitigating the memory wall and maximizing data parallelism. Representative optimization techniques, such as tiling~\cite{wolfe1989more, song1999new, jin2001increasing, strzodka2010cache, bondhugula2016diamond} and layout transformation~\cite{henretty2011data, henretty2013stencil}, are widely employed to enhance data locality. To exploit SIMD capabilities (e.g., AVX, SSE), extensive research has been conducted on both manual~\cite{kennedy2001optimizing, wolfe1995high, rawat2018associative} and compiler-assisted vectorization~\cite{li2021automatic,sun2023adaptive}, aiming to streamline instruction alignment. In the context of GPUs, strategies shift toward hiding memory latency within complex hierarchies. These include hierarchy-aware tiling~\cite{nguyen20103, holewinski2012high, grosser2013split,verdoolaege2013polyhedral, maruyama2014optimizing}, temporal blocking via register rotation~\cite{falch2014register}, and optimized prefetching and streaming techniques~\cite{rawat2018domain, zhao2019exploiting, rawat2019optimizing, gysi2021domain}. However, these approaches are largely bounded by the scalar or vector processing units of traditional architectures.

\textbf{Stencil Computation on Tensor Cores.} 
Stencil computation has emerged as a critical frontier within this broadening landscape of Tensor Core applications.
TCStencil~\cite{TCStencil} pioneers the stencil-to-GEMM paradigm, reformulating stencil computations as matrix multiplications to unlock Tensor Core throughput. Subsequent works, such as ConvStencil~\cite{chen2024convstencil} and LoRAStencil~\cite{zhang2024lorastencil}, refine this transformation to minimize memory overheads. Despite these advances, a fundamental limitation persists: adapting stencils to Tensor Cores inevitably incurs computational redundancy. 
To address this, FlashFFTStencil~\cite{han2025flashfftstencil} leverages FFT to maximize arithmetic intensity. Meanwhile, sparsity-centric approaches like SPIDER~\cite{SPIDER} and SparStencil~\cite{li2025sparstencil} employ novel stride-swapping and compilation techniques to align stencil patterns with hardware requirements, exploiting the hardware-intrinsic 2:4 sparsity of Sparse Tensor Cores.

\section{Conclusion}

This paper resolves the apparent contradiction between the memory-bound nature of stencil computations and their observed acceleration on compute-centric Tensor Cores. By quantifying the algorithmic redundancy and arithmetic intensity shifts, we propose an enhanced performance model that rigorously delineates the \textit{sweet spot} where the benefits of high-throughput units outweigh the costs of adaptation. Our analysis reveals that Tensor Cores become viable when temporal fusion effectively shifts the bottleneck. Extensive evaluations across SOTA implementations like ConvStencil and SPIDER corroborate our model, grounding our analysis in empirical results. Consequently, our work reconciles the discrepancy between conventional assumptions and empirical reality, providing a systematic guideline for stencil acceleration.

\bibliographystyle{ACM-Reference-Format}
\bibliography{reference}

\end{document}